\documentclass[10pt]{article} 

\usepackage[accepted]{tmlr}

\usepackage{amsmath,amsfonts,bm}

\def\eqref#1{equation~\ref{#1}}

\def\1{\bm{1}}

\DeclareMathAlphabet{\mathsfit}{\encodingdefault}{\sfdefault}{m}{sl}
\SetMathAlphabet{\mathsfit}{bold}{\encodingdefault}{\sfdefault}{bx}{n}

\usepackage[T1]{fontenc}
\usepackage[utf8]{inputenc}
\usepackage{url}
\usepackage[colorlinks,linkcolor=red,citecolor=blue,urlcolor=blue]{hyperref}
\usepackage{microtype}
\usepackage{xspace}

\usepackage{amsmath}
\usepackage{amssymb}
\usepackage{amsfonts}
\usepackage{nicefrac}

\usepackage{graphicx}
\usepackage{booktabs}
\usepackage{array}
\usepackage{colortbl}
\usepackage{arydshln}
\usepackage{wrapfig}
\usepackage{caption}
\usepackage{subcaption}
\usepackage{makecell}
\usepackage{multirow}
\usepackage{tabularx}

\usepackage{algorithm}
\usepackage{algpseudocode}

\usepackage{xcolor}
\usepackage[most]{tcolorbox}
\usepackage{pgfplots}
\pgfplotsset{compat=1.17}

\usepackage{enumitem}
\usepackage{pifont}
\usepackage[capitalize,noabbrev]{cleveref}

\usepackage{titlesec}
\titlespacing\section{0pt}{4pt plus 2pt minus 2pt}{4pt plus 2pt minus 2pt}
\titlespacing\subsection{0pt}{2pt plus 2pt minus 2pt}{2pt plus 2pt minus 2pt}

\definecolor{royalblue(web)}{rgb}{0.25, 0.41, 0.88}
\definecolor{blue-violet}{rgb}{0.54, 0.17, 0.89}
\definecolor{brightmaroon}{rgb}{0.76, 0.13, 0.28}
\definecolor{darkmagenta}{rgb}{0.55, 0.0, 0.55}
\definecolor{bleudefrance}{rgb}{0.19, 0.55, 0.91}
\definecolor{palatinateblue}{rgb}{0.15, 0.23, 0.89}
\definecolor{whitesmoke}{rgb}{0.96, 0.96, 0.96}
\definecolor{thulianpink}{rgb}{0.87, 0.44, 0.63}
\definecolor{amber(sae/ece)}{rgb}{1.0, 0.49, 0.0}
\definecolor{darkblue}{rgb}{0.0, 0.0, 0.55}
\definecolor{alizarin}{rgb}{0.82, 0.1, 0.26}
\definecolor{asparagus}{rgb}{0.53, 0.66, 0.42}
\definecolor{darkspringgreen}{rgb}{0.09, 0.45, 0.27}
\definecolor{columbiablue}{rgb}{0.61, 0.87, 1.0}
\definecolor{wildblueyonder}{rgb}{0.64, 0.68, 0.82}
\definecolor{trolleygrey}{rgb}{0.5, 0.5, 0.5}
\definecolor{paleaqua}{rgb}{0.74, 0.83, 0.9}
\definecolor{bubblegum}{rgb}{0.99, 0.76, 0.8}
\definecolor{coralred}{rgb}{1.0, 0.25, 0.25}
\definecolor{green(ryb)}{rgb}{0.4, 0.69, 0.2}
\definecolor{flame}{rgb}{0.89, 0.35, 0.13}
\definecolor{bittersweet}{rgb}{1.0, 0.44, 0.37}
\definecolor{darksalmon}{rgb}{0.91, 0.59, 0.48}
\definecolor{emerald}{rgb}{0.31, 0.78, 0.47}
\definecolor{green(pigment)}{rgb}{0.0, 0.65, 0.31}
\definecolor{codegreen}{rgb}{0,0.6,0}
\definecolor{codegray}{rgb}{0.5,0.5,0.5}
\definecolor{codepurple}{rgb}{0.58,0,0.82}
\definecolor{backcolour}{rgb}{0.96,0.96,0.94}
\definecolor{forestgreen}{RGB}{34,139,34}
\definecolor{custompurple}{HTML}{800080}
\definecolor{mypurple}{RGB}{120,81,169}
\definecolor{darkorange}{RGB}{205,102,0}

\tcbuselibrary{listingsutf8}
\newtcblisting{promptbox}{
  colback=gray!5!white,
  colframe=gray!75!black,
  listing only,
  listing options={
    basicstyle=\ttfamily\footnotesize,
    breaklines=true,
    columns=fullflexible,
    keepspaces=true,
    showstringspaces=false
  },
  left=5pt,
  right=5pt,
  top=5pt,
  bottom=5pt,
  sharp corners,
  boxrule=0.5pt
}

\newcommand{\xhdr}[1]{{\noindent\bfseries #1}.}

\newcommand{\method}{\texttt{LRanker}\xspace}

\newcommand{\revise}[1]{\textcolor{black}{#1}}
\newcommand{\rev}[1]{#1}

\title{LRanker: Scaling LLM Ranking to Millions of Candidates with Test-Time Scaling}

\author{\name Tao Feng\thanks{Equal contribution.} \email taofeng2@illinois.edu \\
      \addr University of Illinois Urbana-Champaign
      \AND
      \name Zijie Lei\footnotemark[1] \email zijielei303@meta.com \\
      \addr Meta Monetization AI
      \AND
      \name Zhigang Hua \email zhua@meta.com \\
      \addr Meta Monetization AI
      \AND
      \name Yan Xie \email yanxie@meta.com \\
      \addr Meta Monetization AI
      \AND
      \name Shuang Yang \email shuangyang@meta.com \\
      \addr Meta Monetization AI
      \AND
      \name Ge Liu \email geliu@illinois.edu \\
      \addr University of Illinois Urbana-Champaign
      \AND
      \name Jiaxuan You \email jiaxuan@illinois.edu \\
      \addr University of Illinois Urbana-Champaign}

\def\month{10}
\def\year{2026}
\def\openreview{\url{https://openreview.net/forum?id=9VBpOwbXrZ}}

\begin{document}

\maketitle

\begin{abstract}
Large language models (LLMs) offer strong semantic representations for ranking, but processing candidate texts within limited contexts and repeatedly invoking an LLM for candidate-level comparisons can be costly when candidate pools contain millions of items. We propose \method, a framework for scaling LLM ranking to millions of candidates with test-time scaling. The framework combines two complementary components. A candidate aggregation encoder summarizes offline candidate embeddings using $K$-means clustering and conditions the LLM on compact global candidate information; ranking then uses inner-product scores with cached candidate embeddings. A graph-based test-time scaling mechanism allocates additional inference computation to partition-conditioned query embeddings and ensembles their ranking scores. This separates the ability to score large candidate pools from the use of additional test-time computation to improve ranking quality. We evaluate \method on seven tasks across three scenarios in \textit{RBench}, including a recommendation task with more than \textit{6.8M} candidates. Against the strongest baseline reported for each task, the relative MRR improvements in \textit{RBench-Large} are 27.5\%, 3.9\%, 0.9\%, and 8.9\% on Rec-Movie, Rec-Toy, MS MARCO, and ESCI, respectively. On \textit{RBench-Small}, the reported task scores improve by 26.7\% on Rec-Music and 6.1\% on Routing-Balance; on the 6.8M-candidate Rec-Clothing task, the improvement over Tiger is 27.0\%. Ablations support the complementary contributions of candidate aggregation and test-time ensembling. These results demonstrate a practical approach to million-scale LLM ranking, with test-time computation providing an additional means of improving ranking effectiveness.
\end{abstract}

\section{Introduction}

Using large language models (LLMs) for ranking has already demonstrated remarkable potential \citep{li2023prompt,lin2024data,jiang2023llm}, showing strong capabilities in capturing semantic relevance, adapting to diverse domains, and achieving competitive performance compared to traditional retrieval and ranking methods. However, constraints such as limited context length \citep{rashid2024ecorank,liu2024sliding} and prohibitive computational costs \citep{chen2025tourrank} constrain the candidate scale of many token-space LLM ranking methods, limiting their applicability to real-world scenarios like search and recommendation, where candidate pools often scale to millions. Therefore, our paper aims to raise attention to this pressing research question: \textit{How can we scale LLM ranking to millions of candidates while using additional test-time computation to improve ranking quality?} \rev{We focus on \emph{direct} full-corpus ranking rather than the conventional retrieve-then-rerank cascade. A hard first-stage retrieval cut can drop relevant candidates before reranking, and the reranker sees only the shortlist; \method\ instead conditions on a compact summary of the whole candidate pool, with a prompt length that does not grow with the number of candidates. We discuss the relation to cascades in Section~\ref{sec:related}, but we do not evaluate an end-to-end cascade.}


Existing LLM-based rankers can be broadly distinguished by their input and output formats as shown in Table \ref{Tab:intro}. In terms of input, prior approaches typically adopt one of four strategies: (1) query only \citep{li2023gpt4rec}, (2) query combined with a single candidate \citep{ma2024fine}, (3) query–candidate pairs \citep{qin2023large}, or (4) the full candidate list \citep{pradeep2023rankvicuna,feng2025iranker,sun2023chatgpt}. While the last option quickly becomes infeasible due to the limited context length of LLMs, the first three do not explicitly condition on a summary of the full candidate pool and can therefore miss useful global candidate information. On the output side, most methods directly generate ranking results in the token space, which couples ranking quality with the LLM’s decoding latency and restricts scalability.

Based on the above discussion, we argue that \textit{a promising approach to million-scale LLM ranking is to combine compact global candidate information with embedding-based scoring.}  Nevertheless,  constructing such LLM rankers faces two key challenges. First, when the number of candidates is large, the limited context length of LLMs makes it difficult to model the global candidate information, which can lead to ranking inaccuracies. Second, relying on a single embedding to rank all candidates limits the expressive capacity of the model, thereby constraining its overall potential.

To address these challenges, we propose \method, a framework for scaling LLM ranking to millions of candidates with test-time scaling. Its two components serve complementary roles. First, a candidate aggregation encoder clusters offline candidate embeddings using $K$-means and projects the resulting centroids into a compact soft prompt. Together with inner-product scoring over cached candidate embeddings, this design avoids placing every candidate's text in the LLM context. Second, graph-based test-time scaling forms query embeddings conditioned on different candidate partitions and iteratively refines them through partitioning, pruning, and aggregation. The final ranking ensembles the scores from these query embeddings. Here, test-time scaling means allocating additional inference computation to query representation refinement, rather than increasing model size or generating longer reasoning traces. Candidate aggregation and cached scoring support large candidate pools, while the test-time ensemble provides an additional mechanism for improving ranking quality. For fixed centroid count, width, and depth, prompt length and the number of LLM forward passes do not grow with candidate count; total scoring, sorting, and candidate-processing costs still do.

We evaluate \method on seven tasks across three scenarios in \textit{RBench}. Against the strongest baseline reported for each task, the relative MRR improvements in \textit{RBench-Large} are 27.5\%, 3.9\%, 0.9\%, and 8.9\% on Rec-Movie, Rec-Toy, MS MARCO, and ESCI, respectively. On \textit{RBench-Small}, the reported task scores improve by 26.7\% on Rec-Music and 6.1\% on Routing-Balance; on the 6.8M-candidate Rec-Clothing task, the improvement over Tiger is 27.0\%. These percentages are relative improvements, computed as $(\text{LRanker}-\text{baseline})/\text{baseline}$ using the rounded scores in Tables~\ref{tab:SplitPerformance} and~\ref{tab:domain_specific}. The main million-scale evaluation is on Rec-Clothing. Component ablations isolate the effect of test-time ensembling on the four RBench-Large tasks; Table~\ref{tab:width_depth} separately reports its latency overhead at validation-selected configurations.

\begin{table}[t]
    \vspace{-4mm}
    \caption{\textbf{Comparison of \method with existing LLM-based rankers across four dimensions: input, output, ranking latency, and maximum candidate scale.} Unlike prior approaches, \method leverages aggregated candidate centroids within an efficient LLM-based ranking architecture, \rev{making the length of its input prompt independent of the number of candidates for a fixed number of centroids $K$ (total computation still grows with the number of candidates; see Appendix~\ref{app:latency})} and enabling efficient processing of the information of large-scale candidate sets.}
    \label{Tab:intro}
    \centering
    \setlength{\tabcolsep}{10pt}
    \resizebox{\textwidth}{!}{\begin{tabular}{lccccc}
        \toprule
        \textbf{LLM-based Ranker} & \textbf{Input} & \textbf{Output} & \textbf{Ranking Latency} & \textbf{Largest Candidate Scale} \\
        \midrule
        PRP \citep{qin2023large} & Query+ Candidate Pair & Token  & High  & 100 \\
         RankGPT  \citep{sun2023chatgpt} & Query + Candidate List & Token  & Moderate  & 100 \\
        IRanker \citep{feng2025iranker} & Query + Partial Candidate List & Token & Moderate & 20 \\
        RankLLaMA \citep{ma2024fine} & Query+Single Candidate & Embedding & High  & 200 \\
        \midrule
        \rowcolor{cyan!10} \method & Query + Aggregated Candidate info & Embedding & \textbf{Low} & \textbf{6.81M} \\
        \bottomrule
    \end{tabular}}
\end{table}

\section{Problem Formulation} 


Given a query $q$, the objective of a ranking task \citep{liu2009learning,li2011short,cao2007learning} is to train a ranker $f$ that orders a candidate set $D = \{c_1, c_2, \ldots, c_n\}$ of size $n$. Typically, $D$ can be separated into a positive subset $D_p$ (items that the user truly interacted with, e.g., products actually purchased) and a negative subset $D_n$ (items not chosen). To assess how accurately the ranker retrieves the positives, its performance is evaluated with ranking metrics $E$, such as Normalized Discounted Cumulative Gain (nDCG) \citep{jarvelin2002cumulated} or Mean Reciprocal Rank (MRR) \citep{voorhees1999trec,cremonesi2010performance}.  

Formally, a ranker $\pi$ maps the pair $(q, D)$ into an ordered sequence
\begin{equation}
\pi: (q, D) \mapsto O = \{c_1^{r_1}, c_2^{r_2}, \ldots, c_n^{r_n}\}, \quad O \in \mathbb{S}_n,
\end{equation}
where $r_i$ denotes the position assigned to candidate $c_i$, and $\mathbb{S}_n$ is the space of all permutations over $n$ elements. The learning objective is then to identify the optimal ranker $\pi^*$ within a hypothesis class $\mathcal{F}$ that maximizes the expected evaluation score under the data distribution $\mathcal{Z}$:
\begin{equation}
\pi^* = \arg\max_{f \in \mathcal{F}} \; \mathbb{E}_{(q, D) \sim \mathcal{Z}} \left[ E(\pi(q, D)) \right].
\end{equation}

\section{\method: Million-Scale Ranking with Test-Time Scaling}

\begin{figure}[t]
    \centering
    \vspace{-3mm}
    \includegraphics[width=\linewidth]{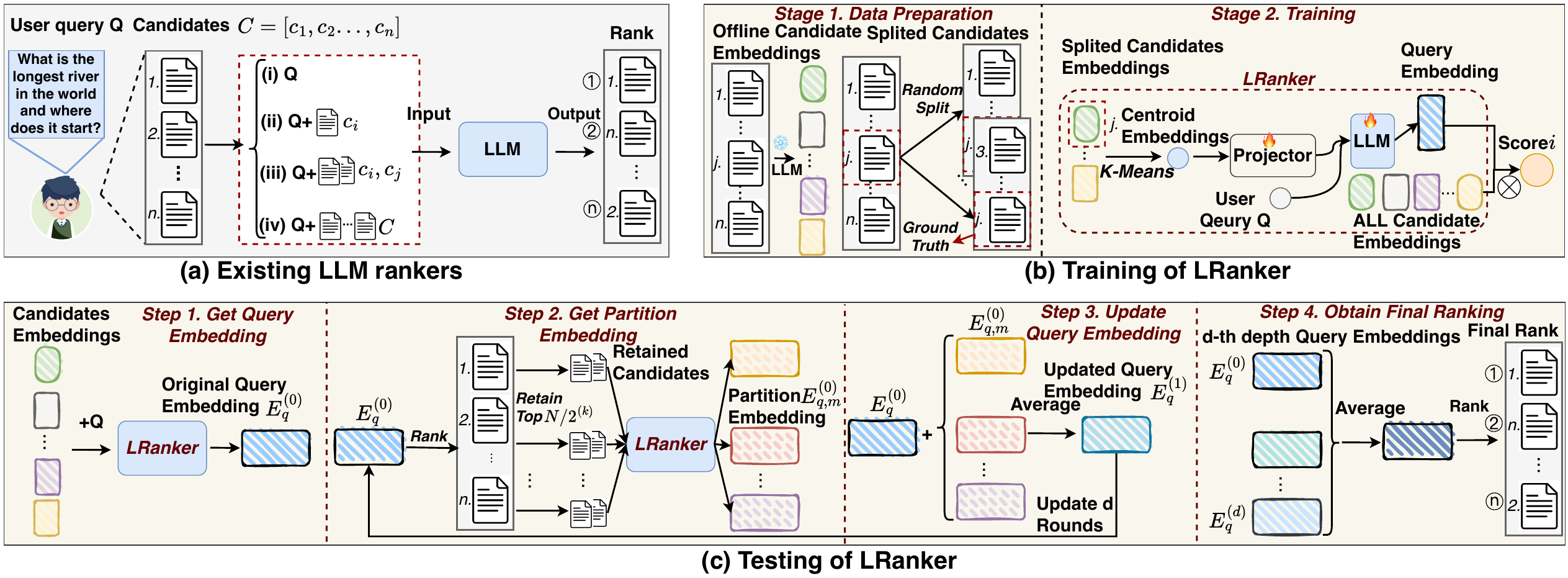}
    \vspace{-0mm}
    \caption{\revise{\textbf{Compared with existing LLM rankers on large-candidate tasks, \method incorporates advanced designs in both the representation of candidate information and the inference strategies used during testing. Note that the spark icon denotes models that require fine-tuning, while the snowflake icon denotes models with frozen weights.} 
    (a) Existing LLM rankers generally adopt four input formats (highlighted in the red box): (i) query only, (ii) query with a single candidate, (iii) query with candidate pairs, and (iv) query with the complete candidate list. The (iv) setting is fundamentally constrained by the limited context length of LLMs in massive-candidate scenarios, while the first three fail to incorporate global candidate-level information, leading to systematic biases in ranking.
    (b) \method’s training pipeline. In Stage~1 (data preparation), offline candidate embeddings are randomly split and clustered by $K$-means to obtain centroid embeddings. In Stage~2 (training), these centroids are passed through the learnable projector and injected into the LLM decoder via a special placeholder token, yielding query embeddings that already condition on global candidate information and are optimized with a ranking loss.
    (c) Graph-based test-time scaling. Step~1 obtains an initial query embedding $E_q^{(0)}$ on the full candidate set. Step~2 uses $E_q^{(0)}$ to retain top-ranked candidates within each partition and recompute partition-specific embeddings. Step~3 updates the query embedding by averaging embeddings from multiple partitions and repeating this elimination-and-update process for $d$ rounds, producing $\{E_q^{(t)}\}_{t=0}^{d}$. Step~4 averages the scores computed from these embeddings to obtain the final ranking, enabling robust performance across diverse candidate scales.}}
    \label{fig:3}
\end{figure}

\revise{Building on the limitations of existing LLM rankers summarized in Figure~\ref{fig:3}, we introduce \method, an encoder--decoder framework designed to handle large-candidate ranking more effectively. \method\ improves ranking performance through two complementary components. First, a \emph{candidate aggregation encoder} applies $K$-means clustering to offline candidate embeddings and uses a learnable projector to inject the resulting aggregated vector into the LLM as a soft prompt, enabling the query representation to condition on global candidate information while candidate representations are cached independently (Figure~\ref{fig:3}(b)). Second, a \emph{graph-based test-time scaling} strategy iteratively partitions and prunes the candidate set, recomputes partition-specific query embeddings, and aggregates them across multiple candidate scales to enhance inference robustness (Figure~\ref{fig:3}(c)). We also describe the training procedure, including random partition sampling for robust query representations, and provide a motivating example to illustrate how these components work together.}

\subsection{Framework of \method}
\label{sec:framework}

We design \method\ as an encoder--decoder framework consisting of two trainable components: a \textit{candidate aggregation encoder} and an \textit{LLM decoder}. The overall goal is to learn query-aware embeddings that integrate both user intent and global candidate information, thereby enabling more accurate passage ranking. Formally, let $\mathcal{P}_\phi$ denote the projection network (with parameters $\phi$) and let $f_\theta$ denote the LLM decoder (with parameters $\theta$). Both $\phi$ and $\theta$ are optimized under the ranking loss described in Section~\ref{sec:training}. The overall pipeline corresponds to Stages~1 and~2 in Figure~\ref{fig:3}(b).

\xhdr{Candidate Aggregation Encoder} 
Following Stage~1 (Data Preparation) in Figure~\ref{fig:3}(b), we first construct a compact representation of the global candidate set. Given a candidate set $\mathcal{C}=\{c_1, c_2, \dots, c_N\}$ of size $N$, we obtain offline base encoder embeddings $\mathbf{e}(c_i)$ for each candidate $c_i$ and cache them for reuse during training and testing. To capture the global distributional structure of candidates, we then apply $K$-means clustering on these embeddings:
\begin{equation}
    \{ \mathcal{G}_1, \mathcal{G}_2, \dots, \mathcal{G}_K \} = \text{KMeans}(\{\mathbf{e}(c_1), \dots, \mathbf{e}(c_N)\}),
\end{equation}
where $\mathcal{G}_k$ represents the $k$-th cluster. Each cluster centroid is computed as
\begin{equation}
    \mathbf{g}_k = \frac{1}{|\mathcal{G}_k|} \sum_{c_i \in \mathcal{G}_k} \mathbf{e}(c_i).
\end{equation}
The $K$ centroids are concatenated and then projected into the LLM embedding space by a learnable projector:
\begin{equation}
    \mathbf{g} = [\mathbf{g}_1; \mathbf{g}_2; \dots; \mathbf{g}_K], 
    \quad 
    \tilde{\mathbf{g}} = \mathcal{P}_\phi(\mathbf{g}),
\end{equation}
where $\mathcal{P}_\phi$ is implemented as a small MLP whose parameters $\phi$ are trained jointly with the LLM parameters $\theta$. In Figure~\ref{fig:3}(b), this corresponds to the red-boxed “Centroid Embeddings” and “Projector” modules that transform offline candidate embeddings into a single aggregated vector. \rev{Appendix~\ref{app:agg_theory} gives the motivation for using $k$-means centroids as a compressed summary of the candidate pool; this choice is supported empirically by the aggregation ablation (Appendix~\ref{app:kpare}) and the sensitivity study (Appendix~\ref{app:k_choice}).}

\xhdr{\revise{LLM as a Decoder}} 
\revise{The LLM $f_\theta$ serves as a decoder that jointly encodes the query and the aggregated candidate information. We design an input prompt that integrates the user query $q$ with the projected aggregated candidate embedding $\tilde{\mathbf{g}}$, as shown in Appendix~\ref{app:prompt} and Stage~2 of Figure~\ref{fig:3}(b). Concretely, we construct a discrete token sequence} \revise{the in-context input $(x_1, x_2, \dots, x_L)$} \revise{that includes system instructions, the textual query $q$, and a special placeholder token \texttt{<|embedding|>} at a fixed position $p$. Let $E_{\text{tok}}$ be the LLM token embedding matrix. We first map all tokens to embeddings and then \emph{replace} the embedding at position $p$ by the continuous vector $\tilde{\mathbf{g}}$:}
\begin{equation}
    \revise{
    \mathbf{x}_t = 
    \begin{cases}
      E_{\text{tok}}(x_t), & t \neq p,\\[3pt]
      \tilde{\mathbf{g}}, & t = p.
    \end{cases}
    }
\end{equation}
\revise{The resulting sequence of input embeddings $\mathbf{X}_0 = (\mathbf{x}_1, \dots, \mathbf{x}_L)$ is then fed into the LLM, producing the hidden representations $(\mathbf{z}_1, \dots, \mathbf{z}_L) = f_\theta(\mathbf{X}_0)$,} \revise{where $\mathbf{z}_t$ denotes the final-layer hidden state at position $t$. Operationally, this is equivalent to using a single-token \emph{soft prompt}: the placeholder position is occupied by a continuous vector $\tilde{\mathbf{g}}$ rather than a discrete token embedding, and gradients flow back to both $\mathcal{P}_\phi$ and $f_\theta$. No additional modifications to intermediate layers are required.}

\revise{To define a query embedding, we explicitly append an end-of-sequence token \texttt{<eos>} to the prompt. For clarity, we use \texttt{<eos>} as a generic notation for the model-specific end-of-text token (e.g., the end-of-sequence token in Qwen-3). For a query $q$ with $T_q$ textual tokens, the full input also contains the instruction and injected soft-prompt positions. The pooling operation below uses only the $T_q$ query-token positions and the \texttt{<eos>} position. Let $\mathbf{z}_{q,t}$ denote the hidden state of the $t$-th query token and $\mathbf{z}_{q,\text{eos}}$ the hidden state at the \texttt{<eos>} position returned by $f_\theta$. Because the \texttt{<eos>} position is used by the LLM to predict the next token, we refer to $\mathbf{z}_{q,\text{eos}}$ as the ``next-token'' hidden state in the sequel. The final query embedding is obtained by averaging over all query-token hidden states and this next-token state:}
\begin{equation}
    \revise{\mathbf{h}_q = \frac{1}{T_q+1} \left( \mathbf{z}_{q,\text{eos}} + \sum_{t=1}^{T_q} \mathbf{z}_{q,t} \right).}
\end{equation}

\revise{Similarly, for each candidate $c_i$ with length $|c_i|$, we obtain an off-line candidate embedding by feeding its text (plus an \texttt{<eos>} token) into the same LLM $f_\theta$.\footnote{For efficiency, this step can be precomputed and cached.} Let $\mathbf{z}_{c_i,j}$ and $\mathbf{z}_{c_i,\text{eos}}$ denote the hidden states of the $j$-th candidate token and the \texttt{<eos>} position, respectively. We define}
\begin{equation}
    \revise{\mathbf{h}_{c_i} = \frac{1}{|c_i|+1} \left( \mathbf{z}_{c_i,\text{eos}} + \sum_{j=1}^{|c_i|} \mathbf{z}_{c_i,j} \right).}
\end{equation}
\revise{Note that $\mathbf{h}_q$ and $\mathbf{h}_{c_i}$ are produced by the same LLM $f_\theta$, with the only difference that $\mathbf{h}_q$ additionally conditions on the injected global candidate vector $\tilde{\mathbf{g}}$ at the placeholder position.}

\revise{Finally, we compute relevance scores and rank the candidates by inner product:}
\begin{equation}
    \revise{
    s(q,c_i) = \langle \mathbf{h}_q, \mathbf{h}_{c_i} \rangle, 
    \quad 
    \pi(q) = \text{argsort}\big( \{s(q,c_i)\}_{i=1}^N \big),
    }
\end{equation}
\revise{where $\langle \cdot, \cdot \rangle$ denotes the inner product and $\pi(q)$ represents the ordered sequence of candidates. This encoder--decoder design (visualized in Figure~\ref{fig:3}(b)) makes it explicit that both the projector $\mathcal{P}_\phi$ and the LLM $f_\theta$ are trainable, and clarifies how the continuous aggregated embedding $\tilde{\mathbf{g}}$ is injected into the discrete token sequence to obtain query and candidate representations.}

\begin{table*}[t]
\small
\setlength{\tabcolsep}{3pt}  
\centering
\renewcommand{\arraystretch}{0.95}  
\caption{\textbf{Detailed summarization of tasks used in our ranking experiments with varying candidate scales.} The table lists seven main RBench tasks and two additional cross-domain evaluation datasets (Rec-Video and Rec-Software; Appendix~\ref{app:gene_exp}).}
\label{tab:dataset_all}
\begin{tabular}{cccc}
    \toprule
    \textbf{Dataset} & \textbf{Task} & \textbf{Candidate Size} & \textbf{\# Queries} \\
    \midrule
    \multicolumn{1}{c}{\textbf{RBench-Small}} \\
    \midrule
    Rec-Music        & Recommendation    & 20     & 12,483 \\
    Routing-Balance  & Routing           & 20     & 1,620  \\
    \midrule
    \multicolumn{1}{c}{\textbf{RBench-Large}} \\
    \midrule
    Rec-Movie        & Recommendation    & 3,884  & 4,167  \\
    Rec-Toy          & Recommendation    & 11,925 & 19,413 \\
    \revise{Rec-Video}    & \revise{Recommendation} & \revise{25,600} & \revise{94,800} \\
    \revise{Rec-Software} & \revise{Recommendation} & \revise{17,600} & \revise{146,400} \\
    MS MARCO         & Passage ranking   & 24,697 & 3,038  \\
    ESCI             & Product searching & 4,000  & 3,999  \\
    \midrule
    \multicolumn{1}{c}{\textbf{RBench-Ultra}} \\
    \midrule
    Rec-Clothing     & Recommendation    & 6,805,462 & 185,925 \\
    \bottomrule
\end{tabular}
\end{table*}

\subsection{Training \method}  
\label{sec:training}

The overall training pipeline in Figure~\ref{fig:3}(b) contains two stages. Stage~1 (Data Preparation) constructs offline candidate embeddings and their $K$-means centroids, while Stage~2 (Training) feeds the aggregated vector into the LLM decoder together with the user query. During this stage, we optimize all trainable parameters 
$\Theta = \{\theta, \phi\}$ of the LLM decoder $f_\theta$ and the projector 
$\mathcal{P}_\phi$ under a ranking objective with both positive and negative samples.
For each query $q$, let $c^+$ denote the ground-truth relevant candidate and 
$\mathcal{C}^- = \{c_1^-, \dots, c_M^-\}$ the set of sampled negative candidates.
Given the relevance score $s(q,c; \Theta)$ defined in Eq.~(9), the model is trained
to assign a higher score to $c^+$ than to any negative candidate $c^- \in \mathcal{C}^-$.
Concretely, the loss function is defined as a softmax cross-entropy:
\begin{equation}
\mathcal{L}(q, c^+, \mathcal{C}^-; \Theta)
= - \log \frac{\exp(s(q,c^+; \Theta))}
{\exp(s(q,c^+; \Theta)) + \sum_{c^- \in \mathcal{C}^-} \exp(s(q,c^-; \Theta))},
\end{equation}
where $s(q,c; \Theta)$ is the inner-product score between the query and candidate
embeddings produced by \method.

\paragraph{Random partition sampling.}
In Figure~\ref{fig:3}(b), the “Random Split” module reflects a key design for robust training: \textit{random partition sampling}. To improve the robustness of the aggregated candidate representation and prepare the model for test-time scaling, we introduce this strategy during training. For each training instance with candidate set
$\mathcal{C} = \{c_1, \dots, c_N\}$, we first randomly split $\mathcal{C}$ into 
$M$ disjoint subsets:
\begin{equation}
    \mathcal{C} = \mathcal{C}^{(1)} \cup \dots \cup \mathcal{C}^{(M)}, 
    \quad 
    \mathcal{C}^{(m)} \cap \mathcal{C}^{(m')} = \emptyset \,\,\text{for } m \neq m'.
\end{equation}
At each optimization step, we then uniformly sample one index 
$r \in \{1, \dots, M\}$ and compute the aggregated candidate vector
\begin{equation}
    \tilde{\mathbf{g}}^{(r)} = \mathcal{P}_\phi\big(\text{Aggregate}(\mathcal{C}^{(r)})\big),
\end{equation}
where \text{Aggregate}$(\cdot)$ denotes the $K$-means-based centroid extraction
described in Section~\ref{sec:framework}. The sampled vector 
$\tilde{\mathbf{g}}^{(r)}$ is injected into the prompt via the special placeholder
token (Eqs.~(4)--(5)), and the LLM decoder $f_\theta$ produces the query embedding
$\mathbf{h}_q^{(r)}$ and candidate embeddings. These embeddings are then used to
compute scores $s(q, c; \Theta)$ and update the parameters via the loss
$\mathcal{L}(q, c^+, \mathcal{C}^-; \Theta)$.

This random partition sampling serves as a form of data augmentation over candidate
contexts: \rev{during training, the model sees each query with many different}
\rev{subsets of candidates and candidate scales.} \rev{We emphasize that this partitioning is \emph{random by design} rather than based on semantic, user, or temporal structure: a structured partition would tie every view to the same grouping, whereas random splits are intended to give the ensemble diverse views. We present this as a motivation; we do not claim that random partitioning makes the learned, pruned ensemble an unbiased or variance-reducing estimator. Because the final ranking score is still an inner product computed per candidate (Eq.~(9)), random partitioning only diversifies the contexts in which the query embedding is formed and never severs a query--candidate relationship.} \rev{Random partition sampling is intended to make} the “Query Embedding” output in Figure~\ref{fig:3}(b) \rev{less sensitive to the particular candidate context.} Consequently, at test time,
\method\ can naturally leverage multiple candidate partitions of different sizes and
aggregate their contributions for improved ranking performance, as discussed in
Section~\ref{sec:tts}.

\subsection{Graph-Based Test-Time Scaling}
\label{sec:tts}

Our test-time scaling mechanism allocates additional LLM query encodings to partition-conditioned representations while holding model parameters fixed. Motivated by ensemble learning \citep{zhou2025ensemble,breiman1996bagging,freund1997decision,wolpert1992stacked}, it combines query embeddings and averages their ranking scores across refinement rounds (Figure~\ref{fig:3}(c)). This can improve ranking quality, but does not guarantee monotonic improvements as width or depth increases. With cached candidate embeddings and the linear scoring rule, averaging scores is algebraically equivalent to scoring with the average query embedding; the additional computation changes how that query representation is constructed.

Concretely, for a ranking task with $N$ candidates, we first obtain an initial query embedding $E_q^{(0)}$ using the proposed encoder--decoder framework applied to the full candidate set (Step~1 in Figure~\ref{fig:3}(c)). Based on $E_q^{(0)}$, we perform an \emph{elimination} step (Step~2): the candidates are partitioned into $k$ disjoint subsets of size approximately $N/k$, and within each subset we retain only the top-ranked candidates according to $E_q^{(0)}$. This yields a reduced candidate pool $\mathcal{C}^{(1)}$ that is still diverse but more focused on promising items. \rev{As in training, this partition into $k$ subsets is drawn uniformly at random rather than by semantic, user, or temporal grouping, so that the views are not all tied to one grouping.}

For each of the $k$ subsets in this reduced pool, we recompute the aggregated candidate information (via the same $K$-means-based encoder in Section~\ref{sec:framework}) and obtain a new query embedding conditioned on that subset (Step~3). This yields $k$ embeddings, which are then averaged with the original $E_q^{(0)}$ to produce an updated embedding $E_q^{(1)}$:  
\begin{equation}
E_q^{(1)} = \frac{1}{k+1} \left(E_q^{(0)} + \sum_{m=1}^{k} E_{q,m}^{(0)} \right),
\end{equation}
where $E_{q,m}^{(0)}$ denotes the embedding obtained from the $m$-th partition in the first elimination round. Intuitively, $E_q^{(1)}$ aggregates information from multiple “views” of the candidate set.

\rev{Although the initial centroids are global, the pool retained in Step~2 is selected by the query-conditioned embedding $E_q^{(0)}$; the centroids recomputed over this retained pool in subsequent rounds are therefore \emph{implicitly query-conditioned}, giving \method a lightweight form of query-aware aggregation rather than relying on a single fixed global summary.} This elimination-and-update process can be repeated for multiple iterations, producing a sequence of embeddings $E_q^{(0)}, E_q^{(1)}, \dots, E_q^{(d)}$. At test time (Step~4), the final ranking score for a candidate $c$ is computed by averaging its scores across all embeddings in the sequence:  
\begin{equation}
s_{\text{final}}(q,c) = \frac{1}{d+1} \sum_{t=0}^{d} s_{E_q^{(t)}}(q,c),
\end{equation}
where $s_{E_q^{(t)}}(q,c)$ is the score computed with embedding $E_q^{(t)}$ using the same inner-product function as in Eq.~(9).

\begin{figure}[t]
  \centering
  \includegraphics[width=0.6\textwidth]{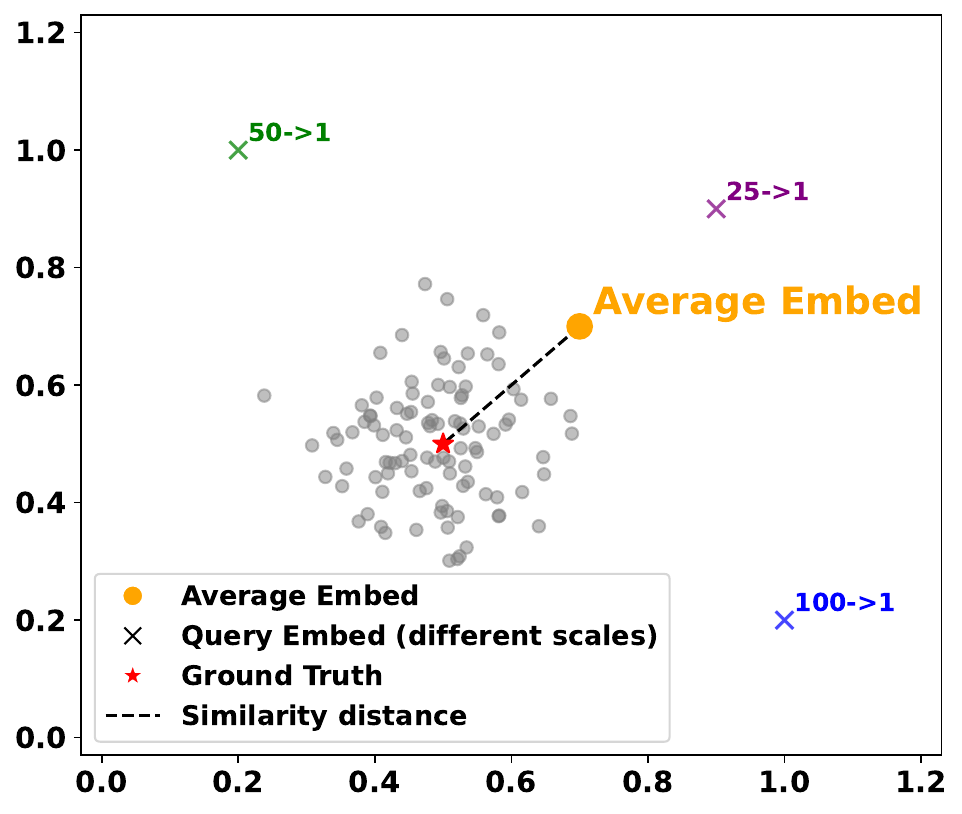}
  \caption{\textbf{Qualitative visualization of partition-conditioned query embeddings and their average.}
  In this example, the average is closer to the ground-truth item in the two-dimensional plot; distances in this visualization do not establish improvements in the original embedding space.}
  \label{fig:qualitative_example}
\end{figure}

We refer to $k$ (the number of partitions per elimination step) as the \textit{width} of test-time scaling, and to $d$ (the number of embedding update iterations) as its \textit{depth}. Together, this forms a graph-based scaling procedure, where embeddings propagate along a graph of candidate partitions to refine query representations iteratively, as illustrated in Figure~\ref{fig:3}(c). \rev{Here ``graph-based'' refers to this induced structure---nodes are the partition-conditioned query embeddings produced at each round and edges are the averaging/propagation operations that combine them into the next embedding---rather than a graph neural network: there is no learned message passing and no explicit adjacency matrix is constructed.} In practice, both width~$k$ and depth~$d$ are selected based on validation performance, and the best hyperparameters are directly applied at inference time. \rev{Test-time scaling adds $d\cdot k$ partition-conditioned query encodings to the initial one ($1 + d\cdot k$ query encodings per query), since each of the $d$ rounds performs $k$ partition encodings. These are logical encoding evaluations, not necessarily separate GPU invocations: partitions may be batched within a round, but the $d$ refinement rounds are sequentially dependent. Each round also prunes candidates and recomputes per-partition centroids online from the cached offline embeddings (no re-encoding), and the final scores use all $d+1$ query embeddings (Eq.~(14)); unlike the number of LLM passes, the cost of these steps grows with the number of candidates. With the validation-selected $k$ and $d$, Table~\ref{tab:width_depth} reports latency increases of 1.2--7.8\% over direct ranking; these reported wall-clock measurements are distinct from the encoding count.}

\subsection{Qualitative Illustration}

We provide a qualitative illustration in Figure \ref{fig:qualitative_example} to show how \method’s test-time scaling mechanism leverages multiple query embeddings to enhance representation quality compared to using a single embedding in a dual-tower setup. On a recommendation dataset with 100 candidates, we apply t-SNE to visualize candidate embeddings, the ground-truth item embedding, and the query embeddings produced at different $N \!\to\! 1$ scales. The plots indicate that the averaged query embedding from test-time ensemble integrates complementary strengths from individual embeddings and qualitatively appears closer to the ground-truth, providing intuitive evidence for its potential to improve ranking performance.

\section{Experiments} \label{sec:exp}

To explore the capability of LLMs for large-candidate ranking, we conduct a comprehensive training and evaluation of the proposed \method across seven interdisciplinary tasks in LLM ranking bench (RBench) with varying candidate scales. We then compare its performance against both general ranking baselines and domain-specific methods. We begin by introducing the tasks within the LLM ranking framework.

\xhdr{Task description}  The details of the tasks are summarized across three scenarios in Table \ref{tab:dataset_all}.

\begin{itemize}[leftmargin=1em, itemsep=0pt]
    \item \textbf{RBench-Large.} For the large-candidate ranking scenario, we employ four widely used datasets, where the number of candidates ranges from several thousand to over ten thousand as shown in Table \ref{tab:dataset_all}. We begin our experiments with two sequential recommendation datasets, MovieLens ml-1m (Rec-Movie) \citep{hou2024large}  and Amazon Toys (Rec-Toy) \citep{mcauley2015image,ni2019justifying}. For both tasks, following prior studies \citep{geng2022recommendation,hua2023index}, we construct each sample by extracting 20 consecutive interactions as the historical sequence, while designating the 21st interaction as the ground-truth item. For evaluation, we adopt the widely used leave-one-out strategy.  In addition, we adopt datasets from passage ranking and product search tasks, namely MS MARCO \citep{nguyen2016ms} and ESCI \citep{reddy2022shopping}, respectively. For both datasets, we assign one positive sample to each query and construct the negative sample set for each query by aggregating the negative samples from all queries. We split the data into training, validation, and test sets using an 8:1:1 ratio.

    \item \textbf{RBench-Ultra.} This scenario is designed to investigate the capability limits of the LLM-based \method in addressing ultra-large ranking tasks with candidate pools at the million scale. To this end, we employ the sequential recommendation dataset Amazon Clothing (Rec-Clothing) \citep{mcauley2015image,ni2019justifying}, which contains nearly 7 million candidate items as shown in Table \ref{tab:dataset_all}. For this task, we follow the same setting as in the RBench-Large setup, namely extracting 20 consecutive interactions as the historical sequence and designating the 21st interaction as the ground-truth item. Evaluation is also conducted using the widely adopted leave-one-out strategy.
    
    \item \textbf{RBench-Small.} We design this scenario to explore the capability of \method in ranking scenarios with relatively small candidate sets, such as the re-ranking stage in recommender systems. To this end, following prior work \citep{feng2025iranker, feng2024graphrouter}, we adopt the sequential recommendation task Rec-Music and the LLM routing task Routing-Balance, both with 20 candidates as shown in Table \ref{tab:dataset_all}, which is fully consistent with the setting in \citep{feng2025iranker}.  
    
\end{itemize}

\xhdr{Baselines and metrics} We evaluate a variety of baseline methods across three scenarios. The baselines are categorized into two groups: \textbf{(a) \textit{General baselines}} that apply across tasks, and \textbf{(b) \textit{Task-specific baselines}} tailored to each task.  For all methods, we primarily use Mean Reciprocal Rank (MRR) \citep{voorhees1999trec,cremonesi2010performance} and  Normalized Discounted Cumulative Gain (NDCG@K) \citep{jarvelin2002cumulated,burges2005learning,liu2009learning} with K = 10  to evaluate ranking performance in the main text. 

\begin{itemize}[leftmargin=1em, itemsep=0pt]

\item \textbf{General baselines.} We consider two categories of general baselines: retrieval-based methods and LLM-based methods. In retrieval-based methods, user queries or histories are treated as the query, while candidates are regarded as the corpus. We employ both a classical probabilistic retrieval model and a modern dense retrieval model:  
1) \textit{BM25} \citep{robertson2009probabilistic}, a traditional probabilistic retrieval function that leverages term frequency, inverse document frequency, and document length normalization. 2) \textit{Contriever} \citep{izacard2021unsupervised}, a state-of-the-art dense retrieval model trained with contrastive learning and hard negatives. 


\item \textbf{Task-specific baselines.} For the \textbf{recommendation tasks in scenarios of RBench-Large and RBench-Ultra}, following prior work on large-scale recommendation \citep{rajput2023recommender}, we implemented five sequential recommendation baselines:  
1) \textit{FM} \citep{rendle2010factorization}: A general predictive model that efficiently captures all pairwise feature interactions, widely used as a strong baseline for recommendation and CTR prediction.  
2) \textit{BERT4Rec} \citep{sun2019bert4rec}: A sequential recommender that applies the bidirectional Transformer (BERT) architecture to user–item sequences, enabling effective modeling of complex item dependencies.  
3) \textit{GRU4Rec} \citep{hidasi2015session}: A session-based recommendation model that leverages gated recurrent units (GRUs) to capture sequential dependencies in user interaction data.  
4) \textit{SASRec} \citep{kang2018self}: A Transformer-based sequential recommender that employs self-attention to model both short- and long-term user preferences.  
5) \textit{Tiger} \citep{rajput2023recommender}: A state-of-the-art generative retrieval framework designed for large-scale recommendation, which encodes items into semantic IDs and autoregressively generates them for efficient ranking under massive candidate sets.

For the \textbf{recommendation tasks in RBench-Small scenario}, we follow the baseline setup in \citep{feng2025iranker} and adopt three representative methods:  
1) \textit{SASRec} \citep{kang2018self}: A self-attention-based sequential recommender that models users' sequential behavioral patterns using a Transformer architecture.  
2) \textit{BPR} \citep{rendle2012bpr}: A pairwise ranking method that optimizes sequential recommendation by encouraging observed items to be ranked higher than unobserved ones.  
3) \textit{R1-Rec} \citep{lin2025rec}: A reinforcement learning-based framework that directly optimizes retrieval-augmented LLMs for recommendation tasks using downstream feedback. As for the \textbf{routing task}, we compared three representative routers:  
1) \textit{RouterKNN} \citep{hu2024routerbench}: A simple yet effective routing baseline that assigns queries to models by retrieving similar examples and applying majority voting.  
2) \textit{RouterBERT} \citep{ong2024routellm}: A lightweight BERT model fine-tuned for routing decisions using classification over task labels.  
3) \textit{GraphRouter} \citep{feng2024graphrouter}: A state-of-the-art graph-based router that balances performance and cost through structural modeling.

Finally, for the \textbf{passage ranking and product search tasks}, we implemented five specialized ranking baselines:
1) \textit{RankBERT-110M} \citep{nogueira2019passage}: A BERT-based passage reranker fine-tuned on MS MARCO relevance judgments, treating ranking as a binary classification problem.  
2) \textit{Multilingual-E5-560M} \citep{wang2022text}: A multilingual embedding model optimized for retrieval and ranking tasks, trained with contrastive learning objectives to generate semantically meaningful embeddings across languages. 
3) \textit{KaLM-mini-instruct-0.5B} \citep{hu2025kalm}: A 0.5B-parameter multilingual embedding model, instruction-tuned for retrieval and ranking tasks.
4) \textit{BGE-Rerank-v2-m3-568M} \citep{xiao2024c}: A state-of-the-art reranker from the BGE series, fine-tuned on large-scale relevance datasets to enhance cross-encoder-based ranking performance.
5) \textit{RankLLama-8B} \citep{ma2024fine}: A ranking-specialized version of Llama-2-8B fine-tuned for passage ranking using pairwise and listwise objectives.
\end{itemize}

\rev{We make the intended scale regime of each baseline explicit. The RBench-Small recommendation and routing baselines (e.g., SASRec, BPR, R1-Rec, GraphRouter) and the passage/product reranking baselines (RankBERT, Multilingual-E5, KaLM, BGE-Rerank-v2, RankLLaMA) are designed for small-to-moderate reranking and are accordingly evaluated in the RBench-Small and RBench-Large scenarios; they are not intended as million-scale competitors. Our million-scale evaluation instead compares against recommenders built for large candidate pools: on the 6.8M-candidate Rec-Clothing task (RBench-Ultra, Table~\ref{tab:domain_specific}) we use the sequential recommenders BERT4Rec, GRU4Rec, and SASRec together with the generative-retrieval model Tiger~\citep{rajput2023recommender}, and the scalability study in Appendix~\ref{app:scalability} extends the \method--Tiger comparison to 48M candidates, where \method's MRR degrades by roughly 25\% from 6.8M to 48M versus around 50\% for Tiger.}

\begin{table*}[t]
\setlength\tabcolsep{3pt}
\centering
\begin{footnotesize}
\vspace{-6mm}
\caption{\textbf{Model performance comparison with general ranking baselines and task-specific baselines across four tasks on NDCG@10 and MRR}. Left: Rec-Movie and Rec-Toy. Right: MS MARCO and ESCI. \textbf{Bold} and \underline{underline} denote the best and second-best results. \rev{For a fair comparison, all learnable baselines are fine-tuned on the same per-dataset training splits as \method\ under an identical train/validation/test protocol; no baseline is evaluated zero-shot.}}
\label{tab:SplitPerformance}
\begin{minipage}[t]{0.49\textwidth}
\centering
\resizebox{\linewidth}{!}{
\begin{tabular}{cccccc}
\toprule
& \multicolumn{2}{c}{\textbf{Rec-Movie}} & \multicolumn{2}{c}{\textbf{Rec-Toy}} \\
\cmidrule(lr){2-3} \cmidrule(lr){4-5}
\textbf{Model} & $\mathrm{NDCG}@10$ & $\mathrm{MRR}$ & $\mathrm{NDCG}@10$ & $\mathrm{MRR}$ \\
\midrule
\multicolumn{1}{c}{\textbf{\textit{General Ranking Baselines}}} \\
\midrule
BM25       & 0.18 & 0.54 & 0.37 & 0.42 \\
Contriever & 0.24 & 0.43 & 0.84 & 1.11 \\
\midrule
\multicolumn{1}{c}{\textbf{\textit{Task-specific Baselines}}} \\
\midrule
FM     & 2.35 & 2.01 & 0.95 & 0.98 \\
BERT4Rec     & 4.08 & 3.56 & 1.26 & 1.31 \\
GRU4Rec     & 4.12 & 3.59 & 1.59 & 1.46 \\
SASRec     & 4.36 & 3.84 & 1.65 & 1.52 \\
Tiger     & \underline{7.37} & \underline{6.12} & \underline{2.99} & \underline{2.33} \\
\midrule
\midrule
\method & \textbf{8.02} & \textbf{7.80} & \textbf{3.21} & \textbf{2.42} \\
\bottomrule
\end{tabular}}
\end{minipage}
\hfill
\begin{minipage}[t]{0.49\textwidth}
\centering
\resizebox{\linewidth}{!}{
\begin{tabular}{ccccc}
\toprule
& \multicolumn{2}{c}{\textbf{MS MARCO}} & \multicolumn{2}{c}{\textbf{ESCI}} \\
\cmidrule(lr){2-3} \cmidrule(lr){4-5}
\textbf{Model} & $\mathrm{NDCG}@10$ & $\mathrm{MRR}$ & $\mathrm{NDCG}@10$ & $\mathrm{MRR}$ \\
\midrule
\multicolumn{1}{c}{\textbf{\textit{General Ranking Baselines}}} \\
\midrule
BM25       & 34.77 & 26.28 & 33.70 & 23.77 \\
Contriever & 44.36 & 33.29 & 29.41 & 26.17    \\
\midrule
\multicolumn{1}{c}{\textbf{\textit{Task-specific Baselines}}} \\
\midrule
RankBERT-110M     & 42.26 & 28.59 & 42.39 & 31.45    \\
Multilingual-E5-560M & \underline{53.73} & 46.49 & 53.17 & 48.43    \\
KaLM-mini-instruct-0.5B & 50.57 & 40.07    & 55.21 & 50.28 \\
BGE-Rerank-v2-m3-568M    & 53.43 & 47.74    & 52.02 & 48.10    \\
RankLLaMA 8B     & 52.22 & \underline{48.83}    & \underline{55.78} & \underline{52.37}    \\

\midrule
\midrule
\method & \textbf{54.80} & \textbf{49.28} & \textbf{58.80} & \textbf{57.01} \\
\bottomrule
\end{tabular}}
\end{minipage}
\end{footnotesize}
\end{table*}

\begin{table}[t]
\centering
\caption{\textbf{Compared with state-of-the-art domain-specific baselines, \method consistently outperforms them across both ultra-long and ultra-short scenarios.} Each task is evaluated with the task-specific metric described in Section~\ref{sec:exp}; values are absolute (\%), with the best in \textbf{bold} and the second best \underline{underlined}. Rec-Music and Routing-Balance belong to the RBench-Small scenario, while Rec-Clothing belongs to the RBench-Ultra scenario. This table reports the same results previously shown as a normalized bar chart; absolute values are given here because the three tasks operate on very different scales.}
\label{tab:domain_specific}
\begin{footnotesize}
\begin{tabular}{llc}
\toprule
\textbf{Task (Scenario)} & \textbf{Method} & \textbf{Performance (\%)} \\
\midrule
\multicolumn{3}{l}{\textit{Rec-Music} --- RBench-Small, 20 candidates} \\
 & SASRec \citep{kang2018self}        & 22.0 \\
 & BPR \citep{rendle2012bpr}          & 23.7 \\
 & R1-Rec \citep{lin2025rec}          & 17.2 \\
 & IRanker \citep{feng2025iranker}    & \underline{29.2} \\
 & \method                            & \textbf{37.0} \\
\midrule
\multicolumn{3}{l}{\textit{Routing-Balance} --- RBench-Small, 20 candidates} \\
 & GraphRouter \citep{feng2024graphrouter} & 21.7 \\
 & RouterBert \citep{ong2024routellm}      & 22.0 \\
 & RouterKNN \citep{hu2024routerbench}     & \underline{29.6} \\
 & IRanker \citep{feng2025iranker}         & 24.4 \\
 & \method                                 & \textbf{31.4} \\
\midrule
\multicolumn{3}{l}{\textit{Rec-Clothing} --- RBench-Ultra, 6.8M candidates} \\
 & BERT4Rec \citep{sun2019bert4rec}     & 0.26 \\
 & GRU4Rec \citep{hidasi2015session}    & 0.25 \\
 & SASRec \citep{kang2018self}          & 0.29 \\
 & Tiger \citep{rajput2023recommender}  & \underline{0.37} \\
 & \method                              & \textbf{0.47} \\
\bottomrule
\end{tabular}
\end{footnotesize}
\end{table}

\subsection{\method Outperforms General Ranking Methods and Task-Specific Baselines}

In the RBench-Large scenario, we compare the \textbf{0.6B-sized} \method across four tasks (Table~\ref{tab:SplitPerformance}). Relative to the strongest baseline in each MRR column, the improvements are 27.5\% on Rec-Movie (7.80 vs.\ 6.12), 3.9\% on Rec-Toy (2.42 vs.\ 2.33), 0.9\% on MS MARCO (49.28 vs.\ 48.83), and 8.9\% on ESCI (57.01 vs.\ 52.37). The corresponding NDCG@10 improvements are 8.8\%, 7.4\%, 2.0\%, and 5.4\%. Thus, the gains differ substantially by task and metric; \method has the highest reported point estimate in each column, but these comparisons do not establish statistical significance.

\subsection{\method Achieves Superior Results in Both RBench-Ultra and RBench-Small Scenarios}

In RBench-Small and RBench-Ultra, we compare the 0.6B-sized \method with the task-specific baselines in Table~\ref{tab:domain_specific}. On Rec-Music, the score increases from 29.2\% for IRanker to 37.0\% for \method: a 7.8-percentage-point increase, or 26.7\% relative improvement. On Routing-Balance, it increases from 29.6\% for RouterKNN to 31.4\%: 1.8 percentage points, or 6.1\% relative improvement. On Rec-Clothing, with 6,805,462 candidates, the score increases from 0.37\% for Tiger to 0.47\%: 0.10 percentage points, or 27.0\% relative improvement. These are full-method comparisons; they do not isolate the contribution of test-time scaling on the million-scale task.

\begin{figure}[t]
    \centering
    \vspace{-4mm}
    \includegraphics[width=0.9\linewidth]{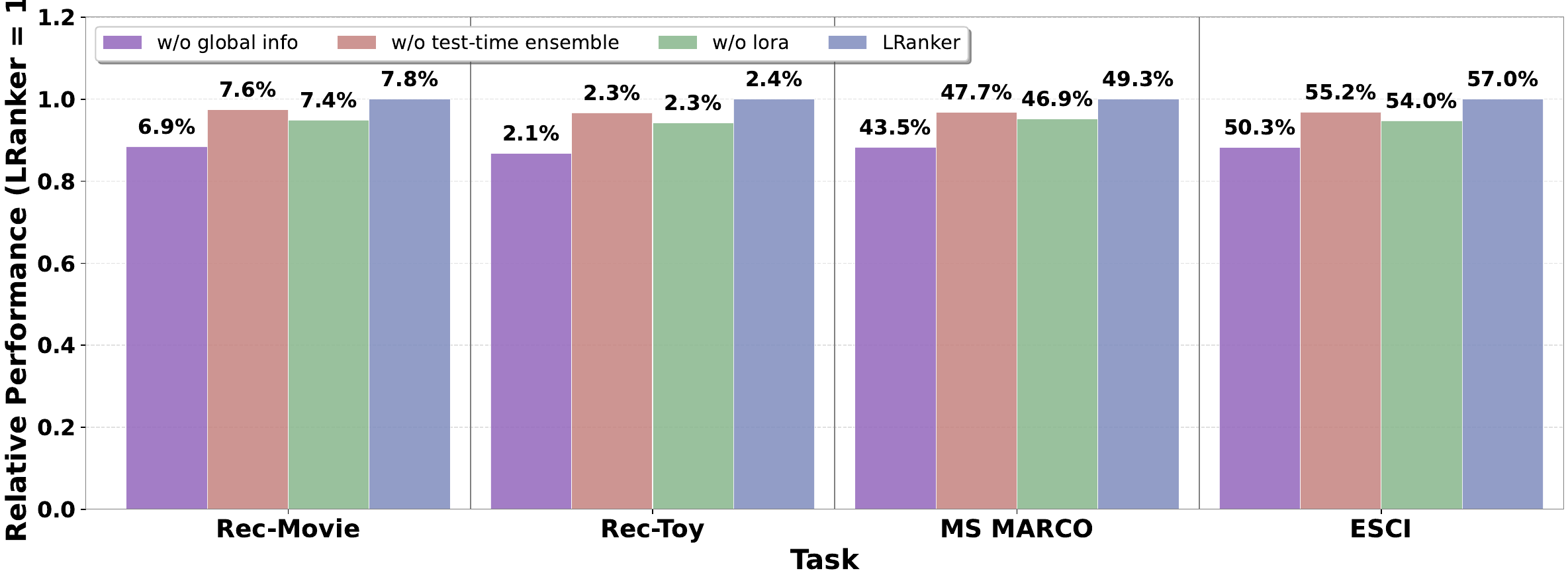}
    \vspace{-0mm}
    \caption{ \textbf{Ablation studies confirm that each component of \method contributes positively to the overall performance.} To further examine their roles, we evaluate three ablated settings: (i) w/o global info removes aggregated candidate information, excluding the clustered embedding input and its projector; (ii) w/o test-time ensemble disables the ensemble mechanism, relying only on the initial embedding from the LLM; and (iii) w/o LoRA freezes LLM parameters during training and only fine-tunes the projector. Bar heights are normalized to the full model; annotations give rounded absolute MRR in percent. As shown across Rec-Movie, Rec-Toy, MS MARCO, and ESCI, removing any component consistently leads to performance degradation.
    }
\label{Fig:abl}
\end{figure}

\subsection{Ablation Studies Confirm the Effectiveness of \method’s Key Components}

To provide a comprehensive understanding of the key components of \method, we conduct a series of experiments to investigate the effect of different components. 

\begin{itemize}[leftmargin=1em, itemsep=0pt]
    \item \textbf{w/o global info}: Evaluates the contribution of incorporating global candidate information. This variant removes the clustered embedding input and its associated projector from the \method framework. 
    
    \item \textbf{w/o test-time ensemble}: Assesses the impact of the test-time ensemble mechanism. In this setting, \method performs ranking solely using the initial embedding generated by the LLM.  
    
    \item \textbf{w/o LoRA}: Examines the role of LoRA-based training. Here, the LLM parameters are frozen during training, and only the projector is fine-tuned.  
\end{itemize}

We report the evaluation results on Rec-Movie, Rec-Toy, MS MARCO, and ESCI datasets in Figure \ref{Fig:abl}. It can be observed that removing global candidate information causes a clear degradation across all tasks. Without clustered embeddings and the corresponding projector, the model fails to capture global context, which limits its ability to discriminate among candidates and lowers ranking accuracy. Removing the test-time ensemble mechanism also leads to reduced performance. Without the ensemble, the model relies solely on a single embedding from the LLM, which weakens its adaptability to task-specific variations and reduces robustness. Nevertheless, the results without this mechanism remain reasonably strong, suggesting that the ensemble mainly serves as a performance booster. This indicates that while test-time ensemble can further improve ranking quality, users who prioritize inference efficiency may choose to omit it with only a modest loss in performance. This highlights the flexibility of LRanker in accommodating different application needs. Eliminating LoRA-based training likewise results in a performance drop. Freezing the LLM parameters and only fine-tuning the projector prevents the model from learning task-specific adaptations, making it harder to fully exploit the LLM’s representational capacity. \rev{Beyond these component ablations, \method is also robust to its main hyperparameters: it outperforms the strongest baseline at every number of centroids $K \in \{10,20,30,40,50\}$ (Appendix~\ref{app:k_choice}, Figure~\ref{fig:k_sensitivity}) and at every centroid dimensionality in $\{256,512,768,1024\}$ (Appendix~\ref{app:dimension}, Figure~\ref{fig:dimension_sensitivity}), while the validation-selected test-time width and depth add below 8\% latency over direct ranking (Appendix~\ref{app:width_depth}, Table~\ref{tab:width_depth}).}

\section{Additional Related Work}\label{sec:related}

Recent works have explored leveraging large language models (LLMs) for ranking under different paradigms. Token-space ranking methods treat LLMs as text rankers by converting queries and candidates into textual prompts, either through iterative elimination (IRanker \citep{feng2025iranker}, PRP \citep{qin2023large}) or one-shot generation (DRanker \citep{feng2025iranker}, RankVicuna \citep{pradeep2023rankvicuna}). However, these approaches face efficiency and context length limitations for large candidate sets. Embedding-based paradigms address this: single-tower methods (RankLLaMA \citep{ma2024fine}) use neural scoring heads but require independent candidate scoring, while dual-tower methods improve efficiency through pre-computed embeddings but limit expressiveness with single query embeddings. Generative LLM approaches have shown strong performance across diverse ranking tasks \citep{liu2024llmemb,sun2023rankgpt,yoon2024listt5,chen2025rankr1,hou2024llmrank}. Prompting-based methods (PRP \citep{qin2023large}, LLM4Rec \citep{hou2024large}) leverage LLM generalization with minimal modification, while instruction tuning approaches (GPT4Rec \citep{li2023gpt4rec}, RankRAG \citep{yu2024rankrag}) fine-tune models for domain-specific ranking signals. These works highlight LLMs' potential as general-purpose rankers while exposing limitations in efficiency, scalability, and complex reasoning.

\rev{\textbf{Relation to two-stage retrieve-then-rerank pipelines.} Production ranking systems often adopt a two-stage cascade: a lightweight first stage retrieves a shortlist and a heavier model reranks it. \method\ targets a different regime, direct full-corpus ranking, and the two designs make different trade-offs. A first-stage cut can drop relevant candidates before reranking, and the reranker sees only the shortlist. \method\ conditions on a summary of the whole pool given by $K$ aggregated centroids, with a prompt length that is independent of the number of candidates for fixed $K$; it still scores every candidate and prunes candidates during test-time scaling (Section~\ref{sec:tts}). The individual components of a standard cascade appear as separately evaluated baselines---dense and sparse retrievers (BM25, Contriever) and strong neural rerankers (RankLLaMA, BGE-Rerank-v2, Multilingual-E5)---against which \method\ ranks competitively or better (Table~\ref{tab:SplitPerformance}). We have not evaluated an end-to-end cascade, so we make no claim about how \method\ compares with one in accuracy or latency.}  

\rev{\textbf{Scope of the evaluation.} All experiments use static benchmark snapshots. We did not evaluate streaming updates, deletions, long-tail or cold-start performance, or an end-to-end retrieve-then-rerank cascade; the discussion of these settings in this section and in Appendices~\ref{app:streaming} and~\ref{app:longtail} describes design considerations, not measured behavior.}
\section{Conclusion}

We presented \method, a framework for scaling LLM ranking to millions of candidates with test-time scaling. Compact candidate aggregation and cached embedding-based scoring support large candidate pools, while partition-conditioned query refinement and score ensembling use additional inference computation to improve ranking quality. Experiments across the three \textit{RBench} scenarios, including the recommendation task with more than 6.8M candidates, support the effectiveness of this combination against the evaluated baselines. Ablations demonstrate the contribution of test-time ensembling alongside candidate aggregation. These findings support million-scale ranking and the utility of additional test-time computation in the evaluated settings; they do not establish a universal compute--performance scaling law or superiority over end-to-end retrieve-then-rerank cascades. Future work will investigate broader ranking tasks and adaptive allocation of test-time computation under explicit latency budgets.

\bibliography{Ref}
\bibliographystyle{tmlr}

\newpage
\appendix

\section{Prompt Usage} \label{app:prompt}

To unify the input format across different tasks, we design prompt templates that integrate the user query with the aggregated candidate information through the special token \texttt{<|embedding|>}. 
These templates guide the model to attend not only to the query but also to the global context of candidate representations. 
Specifically, we construct task-specific templates for four representative tasks: Recommendation (Table~\ref{prompt-template:recommendation}), Routing (Table~\ref{prompt-template:routing}), Passage Ranking (Table~\ref{prompt-template:passage-ranking}), and Product Searching (Table~\ref{prompt-template:product-searching}). 
Each template follows a unified structure but adapts the final instruction to match the objective of the corresponding task.

\begin{table}[h]
\centering
\caption{\textbf{Prompt template for Recommendation task.}}
\begin{tabular}{p{13cm}}
\toprule[1.1pt]
Task: Recommendation \\
\\
Query: [USER QUERY] \\
\\
\texttt{<|embedding|>} Based on the global context information (candidate items) and the query above, recommend the most relevant item. \\
\bottomrule[1.1pt]
\end{tabular}
\label{prompt-template:recommendation}
\end{table}

\begin{table}[h]
\centering
\caption{\textbf{Prompt template for Routing task.}}
\begin{tabular}{p{13cm}}
\toprule[1.1pt]
Task: Routing \\
\\
Query: [USER QUERY] \\
\\
\texttt{<|embedding|>} Based on the global context information (candidate LLMs/agents) and the query above, select the most suitable route or model. \\
\bottomrule[1.1pt]
\end{tabular}
\label{prompt-template:routing}
\end{table}

\begin{table}[h]
\centering
\caption{\textbf{Prompt template for Passage Ranking task.}}
\begin{tabular}{p{13cm}}
\toprule[1.1pt]
Task: Passage Ranking \\
\\
Query: [USER QUERY] \\
\\
\texttt{<|embedding|>} Based on the global context information (candidate passages) and the query above, identify the most relevant passage. \\
\bottomrule[1.1pt]
\end{tabular}
\label{prompt-template:passage-ranking}
\end{table}

\begin{table}[h]
\centering
\caption{\textbf{Prompt template for Product Searching task.}}
\begin{tabular}{p{13cm}}
\toprule[1.1pt]
Task: Product Searching \\
\\
Query: [USER QUERY] \\
\\
\texttt{<|embedding|>} Based on the global context information (candidate products) and the query above, return the product that best matches the search intent. \\
\bottomrule[1.1pt]
\end{tabular}
\label{prompt-template:product-searching}
\end{table}

\section{\revise{Implementation details}} \revise{We implement \method on top of the Qwen3-0.6B Embedding model\footnote{\url{https://huggingface.co/Qwen/Qwen3-0.6B}}
 using LoRA adaptation. Before training, we generate 1024-dimensional offline candidate embeddings for each candidate’s context via Qwen3-0.6B. For all compared baselines, to ensure a fair comparison, we use the same 1024-dimensional offline candidate embeddings as LRanker for their candidate representations. \rev{Moreover, all learnable baselines are fine-tuned on the same per-dataset training splits as \method\ under an identical train/validation/test protocol: an 8:1:1 train/validation/test split for MS~MARCO and ESCI, and the leave-one-out protocol for the recommendation tasks. No baseline is evaluated in a zero-shot manner in Table~\ref{tab:SplitPerformance}.} For each query, we construct 10 random splits of its associated candidate set and obtain the corresponding embeddings. During training, the split candidate embeddings are clustered using k-means implemented via scikit-learn\footnote{\url{https://scikit-learn.org/stable/getting_started.html}}, producing candidate centroid embeddings that serve as global structural features. These centroids are further projected through a Linear–BatchNorm–ReLU block to 1024 dimensions and fused with the base encoder’s textual embedding to form the final query representation. 
Training is performed with InfoNCE loss (temperature = 0.15), contrasting positives against sampled negatives. The model is trained for 15 epochs using the AdamW optimizer ($\beta_1=0.9$, $\beta_2=0.999$, weight decay = 0.01). We use a learning rate of $1\times10^{-4}$ with a 10\% linear warm-up followed by cosine decay, and a batch size of 20. LoRA is applied to both attention and feed-forward layers with rank = 32, $\alpha = 64$, and dropout = 0.1. To further improve efficiency and stability, we enable BF16 training, gradient checkpointing, and gradient clipping (norm = 0.5). For evaluation, we determine the best graph depth and width using the validation set, and fix these configurations when testing on the held-out test set. Note that, to ensure inference efficiency, we restrict the search depth to 0–5 and the search width to 0–10. The detailed graph despth and width settings can be seen in Table \ref{tab:width_depth} of Appendix. Moreover, during training we carefully designed parallelized processing that allows multiple queries and multiple partition-embedding plans to run simultaneously, significantly reducing inference latency.  All experiments are conducted on 1 NVIDIA A6000 GPUs.}

\revise{To improve clustering efficiency under large-scale candidate pools, we adopt two optimizations. First, we use MiniBatchKMeans, which processes the full candidate set in small batches to accelerate convergence.
Second, because Qwen3-Embedding is trained with Matryoshka Representation Learning (MRL) \citep{zhang2025qwen3}, we can obtain most semantic information by truncating its embedding to the first 128 dimensions. We therefore compute cluster assignments using only the truncated 128-dimensional embeddings, and then compute the final centroid embeddings using the full 1024-dimensional vectors. These two strategies significantly accelerate clustering and make \method scalable in real-world large-candidate settings.}

\section{\revise{Generalization Experiments}}\label{app:gene_exp}

\subsection{\revise{Generalization to new datasets}}

\begin{table}[h]
\setlength\tabcolsep{3pt}
\centering
\begin{footnotesize}
\vspace{-4mm}
\caption{\revise{\textbf{Model zero-shot performance comparison with general ranking baselines and task-specific baselines on Video Games and Software}.  Specifically, we evaluate the method trained on the Rec-Toy dataset in RBench in a zero-shot manner on the Video Games and Software datasets. \textbf{Bold} and \underline{underline} denote the best and second-best results.}}
\label{tab:SplitPerformance_Video_Software}
\resizebox{0.8\linewidth}{!}{
\begin{tabular}{cccccc}
\toprule
& \multicolumn{2}{c}{\textbf{Video Games}} & \multicolumn{2}{c}{\textbf{Software}} \\
\cmidrule(lr){2-3} \cmidrule(lr){4-5}
\textbf{Model} & $\mathrm{NDCG}@10$ & $\mathrm{MRR}$ & $\mathrm{NDCG}@10$ & $\mathrm{MRR}$ \\
\midrule
\multicolumn{1}{c}{\textbf{\textit{General Ranking Baselines}}} \\
\midrule
BM25       & 0.39 & 0.36 & 0.56 & 0.51 \\
Contriever & 0.90 & 0.93 & 0.64 & 0.67 \\
\midrule
\multicolumn{1}{c}{\textbf{\textit{Task-specific Baselines}}} \\
\midrule
FM         & 1.03 & 1.09 & 3.15 & 3.77 \\
BERT4Rec   & 1.18 & 1.38 & 2.97 & 2.31 \\
GRU4Rec    & 1.15 & 1.27 & \underline{4.89} & \underline{4.31} \\
SASRec     & 1.21 & 1.40 & 2.83 & 2.56 \\
Tiger      & \underline{1.93} & \underline{2.17} & 4.58 & 3.94 \\
\midrule
\midrule
LRanker    & \textbf{2.31} & \textbf{2.61} & \textbf{5.43} & \textbf{4.86} \\
\bottomrule
\end{tabular}}
\end{footnotesize}
\end{table}

\revise{To evaluate the generalization ability of \method on datasets beyond RBench, we first train all methods on Rec-Toy and then perform zero-shot testing on the Video Games  and Software  datasets \citep{mcauley2015image,ni2019justifying} from amazon (see Table \ref{tab:dataset_all} for dataset details). We report the results in Table~\ref{tab:SplitPerformance_Video_Software}. As shown in the table, \method exhibits clear cross-domain generalization, outperforming both general-ranking and task-specific baselines by substantial margins. On Video Games, \method delivers roughly 20\% improvements over the strongest task-specific baseline and well over 100\% gains compared with general-ranking baselines. On Software, the zero-shot advantage remains positive, with \method surpassing the best task-specific method by 11.0\% in NDCG@10 and 12.8\% in MRR and general-ranking baselines by several-fold. These consistent percentage gains across two unseen domains demonstrate that the ranking patterns learned from Rec-Toy transfer effectively, highlighting the robust zero-shot generalization capability of \method.}

\subsection{\revise{Performance analysis in scenarios with extremely irrelevant candidates}}

\begin{table*}[h]
\setlength\tabcolsep{3pt}
\centering
\begin{footnotesize}
\vspace{-3mm}
\caption{\revise{\textbf{Performance comparison in scenarios with extremely irrelevant candidates across two scenarios on NDCG@10 and MRR}. Left: Rec-Toy. Right: MS MARCO. Specifically, we train the models on the training sets of Rec-Toy and MS MARCO, where the candidates used during training are the original candidates of each dataset. During testing, however, we replace the candidates with a mixed pool that combines all candidates from Rec-Movie, Rec-Toy, MS MARCO, and ESCI. In addition, $\Delta$ performance denotes the relative difference in each metric between the results obtained under the mixed-candidate setting and those obtained under the original candidate set, recomputed from the rounded scores in Table~\ref{tab:SplitPerformance}. \textbf{Bold} and \underline{underline} denote the best and second-best results.}}
\label{tab:irrelevant_cand}
\begin{minipage}[t]{0.49\textwidth} 
\centering
\resizebox{\linewidth}{!}{
\begin{tabular}{ccc}
\toprule
& \multicolumn{2}{c}{\textbf{Rec-Toy}} \\
\cmidrule(lr){2-3}
\textbf{Model} & $\mathrm{NDCG}@10$ & $\mathrm{MRR}$ \\
\midrule
\multicolumn{1}{c}{\textbf{\textit{General Ranking Baselines}}} \\
\midrule
BM25       & 0.02 & 0.17 \\
Contriever & 0.12 & 0.21 \\
\midrule
\multicolumn{1}{c}{\textbf{\textit{Task-specific Baselines}}} \\
\midrule
FM         & 0.37 & 0.34 \\
BERT4Rec   & 0.55 & 0.57 \\
GRU4Rec    & 0.56 & 0.60 \\
SASRec     & 0.78 & 0.64 \\
Tiger      & \underline{2.25} & \underline{1.91} \\
\midrule
LRanker    & \textbf{2.43} & \textbf{2.06} \\
\midrule
\multicolumn{1}{c}{\textbf{\textit{$\Delta$ performance}}} \\
\midrule
$\Delta$ Tiger   & $-24.7\%$ & $-18.0\%$ \\
$\Delta$ LRanker & $-24.3\%$ & $-14.9\%$ \\
\bottomrule
\end{tabular}}
\end{minipage}
\hfill
\begin{minipage}[t]{0.49\textwidth}
\centering
\resizebox{\linewidth}{!}{
\begin{tabular}{ccc}
\toprule
& \multicolumn{2}{c}{\textbf{MS MARCO}} \\
\cmidrule(lr){2-3}
\textbf{Model} & $\mathrm{NDCG}@10$ & $\mathrm{MRR}$ \\
\midrule
\multicolumn{1}{c}{\textbf{\textit{General Ranking Baselines}}} \\
\midrule
BM25       & 30.24 & 24.87 \\
Contriever & 39.21 & 31.09 \\
\midrule
\multicolumn{1}{c}{\textbf{\textit{Task-specific Baselines}}} \\
\midrule
RankBERT-110M           & 36.27 & 28.13 \\
Multilingual-E5-560M    & 45.85 & 41.35 \\
KaLM-mini-instruct-0.5B & 43.19 & 38.36 \\
BGE-Rerank-v2-m3-568M   & 45.83 & 43.98 \\
RankLLaMA 8B            & \underline{46.91} & \underline{44.04} \\
\midrule
LRanker                 & \textbf{49.03} & \textbf{46.40} \\
\midrule
\multicolumn{1}{c}{\textbf{\textit{$\Delta$ performance}}} \\
\midrule
$\Delta$ RankLLaMA 8B & $-10.2\%$ & $-9.8\%$ \\
$\Delta$ LRanker      & $-10.5\%$ & $-5.8\%$ \\
\bottomrule
\end{tabular}}
\end{minipage}
\end{footnotesize}
\end{table*}

\revise{We combine candidates from Rec-Movie, Rec-Toy, MS MARCO, and ESCI, and evaluate models trained on the original Rec-Toy and MS MARCO pools (Table~\ref{tab:irrelevant_cand}). LRanker retains the highest reported NDCG@10 and MRR. Relative to the original-pool scores in Table~\ref{tab:SplitPerformance}, its MRR drops by 14.9\% on Rec-Toy versus 18.0\% for Tiger, and by 5.8\% on MS MARCO versus 9.8\% for RankLLaMA. The NDCG@10 comparison is mixed: the relative drops are 24.3\% versus 24.7\% on Rec-Toy and 10.5\% versus 10.2\% on MS MARCO. Thus, higher absolute performance does not imply the smallest relative degradation in every metric.}

\subsection{\revise{Experiment under candidates distribution shift}}

\begin{table*}[h]
\setlength\tabcolsep{3pt}
\centering
\begin{footnotesize}
\vspace{-3mm}
\caption{\revise{\textbf{Performance comparison in scenarios under candidates distribution shift across two scenarios on NDCG@10 and MRR}. Left: Rec-Toy. Right: MS MARCO. Specifically, we train the models on the training sets of Rec-Toy and MS MARCO, where the candidates used during training are the original candidates of each dataset. During testing, we replace the candidate set of Rec-Toy with a mixed candidate pool constructed from both Rec-Movie and Rec-Toy. Similarly, for MS MARCO and ESCI, we replace each candidate set with mixed candidate pools that combine candidates from MS MARCO and ESCI. In addition, $\Delta$ performance denotes the relative difference in each metric between the results obtained under the mixed-candidate setting and those obtained under the original candidate set, recomputed from the rounded scores in Table~\ref{tab:SplitPerformance}. \textbf{Bold} and \underline{underline} denote the best and second-best results.}}
\label{tab:candidate_shift}
\begin{minipage}[t]{0.49\textwidth}
\centering
\resizebox{\linewidth}{!}{
\begin{tabular}{ccc}
\toprule
& \multicolumn{2}{c}{\textbf{Rec-Toy}} \\
\cmidrule(lr){2-3}
\textbf{Model} & $\mathrm{NDCG}@10$ & $\mathrm{MRR}$ \\
\midrule
\multicolumn{1}{c}{\textbf{\textit{General Ranking Baselines}}} \\
\midrule
BM25       & 0.14 & 0.29 \\
Contriever & 0.54 & 0.52 \\
\midrule
\multicolumn{1}{c}{\textbf{\textit{Task-specific Baselines}}} \\
\midrule
FM         & 0.71 & 0.48 \\
BERT4Rec   & 1.22 & 1.28 \\
GRU4Rec    & 1.26 & 1.43 \\
SASRec     & 1.39 & 1.34 \\
Tiger      & \underline{2.34} & \underline{2.27} \\
\midrule
LRanker    & \textbf{2.54} & \textbf{2.36} \\
\midrule
\multicolumn{1}{c}{\textbf{\textit{$\Delta$ performance}}} \\
\midrule
$\Delta$ Tiger   & $-21.7\%$ & $-2.6\%$ \\
$\Delta$ LRanker & $-20.9\%$ & $-2.5\%$ \\
\bottomrule
\end{tabular}}
\end{minipage}
\hfill
\begin{minipage}[t]{0.49\textwidth}
\centering
\resizebox{\linewidth}{!}{
\begin{tabular}{ccc}
\toprule
& \multicolumn{2}{c}{\textbf{MS MARCO}} \\
\cmidrule(lr){2-3}
\textbf{Model} & $\mathrm{NDCG}@10$ & $\mathrm{MRR}$ \\
\midrule
\multicolumn{1}{c}{\textbf{\textit{General Ranking Baselines}}} \\
\midrule
BM25       & 32.61 & 27.22 \\
Contriever & 40.68 & 33.74 \\
\midrule
\multicolumn{1}{c}{\textbf{\textit{Task-specific Baselines}}} \\
\midrule
RankBERT-110M           & 38.52 & 29.59 \\
Multilingual-E5-560M    & \underline{47.95} & 45.21 \\
KaLM-mini-instruct-0.5B & 45.73 & 39.48 \\
BGE-Rerank-v2-m3-568M   & 47.06 & 46.59 \\
RankLLaMA 8B            & 47.08 & \underline{47.27} \\
\midrule
LRanker                 & \textbf{51.41} & \textbf{49.01} \\
\midrule
\multicolumn{1}{c}{\textbf{\textit{$\Delta$ performance}}} \\
\midrule
$\Delta$ RankLLaMA 8B & $-9.8\%$ & $-3.2\%$ \\
$\Delta$ LRanker      & $-6.2\%$ & $-0.55\%$ \\
\bottomrule
\end{tabular}}
\end{minipage}
\end{footnotesize}
\end{table*}

\revise{To evaluate the robustness of \method under candidate distribution shift, we construct two separate mixed candidate pools: one combining candidates from Rec-Movie and Rec-Toy, and the other combining candidates from MS MARCO and ESCI. We then compare all methods that are trained on the original candidate sets of Rec-Toy and MS MARCO but tested on their corresponding mixed candidate pools. The results are shown in Table~\ref{tab:candidate_shift}. We can observe that most baselines experience substantial performance degradation when evaluated on the mixed candidate pools, indicating their limited robustness to candidate distribution shift. In contrast, \method consistently achieves the highest accuracy under both Rec-Toy and MS~MARCO settings and exhibits a significantly smaller performance drop compared to strong task-specific baselines such as Tiger and RankLLaMA 8B. These results demonstrate that \method effectively leverages global candidate information and maintains stable ranking behavior even when the candidate distribution changes at test time.}

\subsection{\rev{Design Considerations for Dynamic Corpora}} \label{app:streaming}

\rev{We did not evaluate \method\ on streaming corpora; this subsection describes design properties that are relevant when candidates are added or updated over time, not measured behavior. First, candidate embeddings $\mathbf{e}(c_i)$ are produced by the frozen base encoder and cached per item (Section~\ref{sec:framework}), so in principle adding or updating an item only requires encoding that single item and leaves every other cached embedding untouched. Second, the global summary is computed with MiniBatchKMeans (Appendix~B), whose centroids can in principle be refreshed incrementally from new mini-batches of item embeddings rather than by re-clustering the entire pool, so the cost of a refresh scales with the number of new items rather than with the total corpus size.}

\rev{Random partition sampling during training (Section~\ref{sec:training}) exposes the projector $\mathcal{P}_\phi$ and decoder $f_\theta$ to many different subsets of a fixed candidate pool, which may help when the pool changes; however, subsets of a fixed pool are not the same as a drifting corpus. The closest evidence we have comes from static test-time changes to the candidate pool---mixed irrelevant candidates (Table~\ref{tab:irrelevant_cand}) and candidate-distribution shift (Table~\ref{tab:candidate_shift})---not from streaming updates. Streaming operation, deletions, periodic re-clustering, and whether the projector needs retraining as the centroids drift remain unevaluated.}

\subsection{\rev{Behavior on Long-Tail and Cold-Start Candidates}} \label{app:longtail}

\rev{A natural question is whether summarizing the candidate pool into $k$ centroids biases \method\ toward dense, popular regions at the expense of sparse, long-tail, or cold-start items. The aggregation step does not remove candidates: the centroid soft prompt is injected once at a fixed prompt position, every candidate still receives its own score $\langle \mathbf{h}_q, \mathbf{h}_{c_i} \rangle$ (Eq.~(9)), and only the graph-based elimination step (Section~\ref{sec:tts}, Step~2) removes candidates, based on query-conditioned scores. This does not rule out popularity bias: the query embedding is conditioned on centroids that reflect dense regions of the pool, the scorer is learned from interaction data, and elimination can prune a relevant rare item. We have not evaluated performance on long-tail or cold-start items.}

\section{\revise{Experiments on Scalability}} \label{app:scalability}

\begin{figure}[h]
    \centering
    \includegraphics[width=0.9\linewidth]{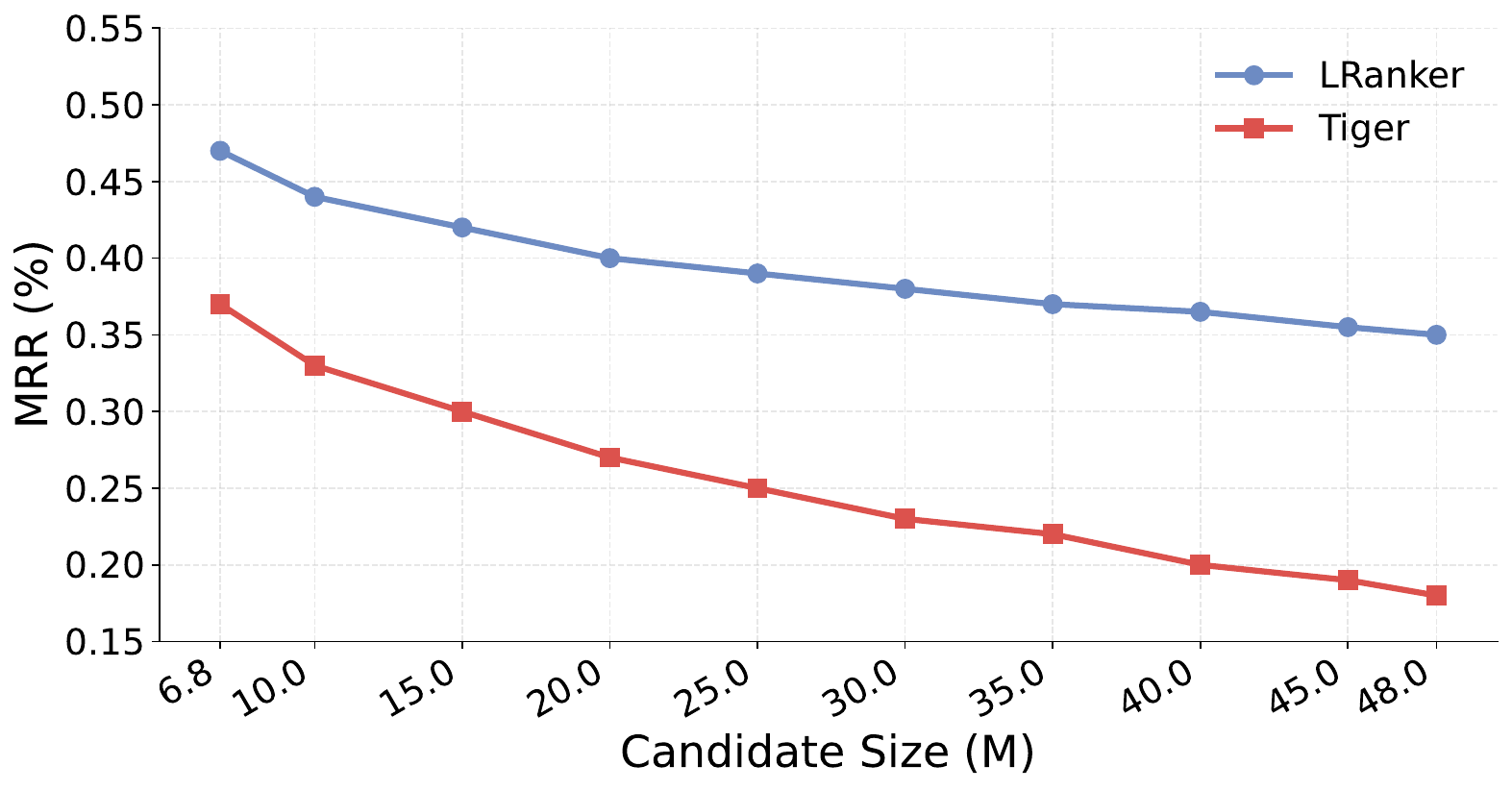}
    \caption{\revise{\textbf{The change in MRR performance of LRanker and Tiger as the candidate size increases.} Candidate pool sizes are 6.8M, 10M, 15M, 20M, 25M, 30M, 35M, 40M, 45M, and 48M.}}
\label{fig:scaling}
\end{figure}

\revise{To examine the limits of LRanker in handling extremely large candidate sets and to analyze how its performance changes under such conditions, we conduct experiments on the Amazon Reviews 2023 dataset \citep{hou2024bridging}, which contains approximately 48.19 million items according to its published dataset statistics. Specifically, we take LRanker and Tiger trained on Rec-Clothing from RBench-Ultra and progressively expand the original 6.8M-candidate pool by randomly adding candidates. We evaluate MRR at candidate pool sizes of 6.8M, 10M, 15M, 20M, 25M, 30M, 35M, 40M, 45M, and 48M. We report the results in Figure~\ref{fig:scaling}. We can observe that both LRanker and Tiger exhibit consistent performance degradation as the candidate size increases, but LRanker maintains a noticeably slower relative decay. From 6.8M to 48M candidates, LRanker’s MRR decreases by roughly \textbf{25\%} relative to its initial value, whereas Tiger suffers a substantially larger relative drop of around \textbf{50\%}. In this experiment, the degradation curves flatten as the pool grows: once the candidate pool grows beyond a certain scale, the decline in ranking performance becomes progressively flatter rather than continuing linearly. This saturation effect likely occurs because, as the candidate pool grows, most newly added items are increasingly irrelevant to the query and therefore less confusable with the ground-truth item. This is a possible explanation rather than a measured property: we do not quantify the number of hard negatives or establish a general scaling law from these curves.}

\section{\revise{Additional Ablation Studies}} \label{app:adb}

\subsection{\revise{Comparative Study of K-Means and Alternative Clustering Techniques}} \label{app:kpare}

\begin{table*}[h]
\setlength\tabcolsep{3pt}
\centering
\begin{footnotesize}
\vspace{-6mm}
\caption{\revise{\textbf{Performance comparison of the possible candidate aggregation encoder variants across four tasks}.}}
\label{tab:aggregation_ablation}

\begin{minipage}[h]{0.49\textwidth}
\centering
\resizebox{\linewidth}{!}{
\begin{tabular}{ccc}
\toprule
& \textbf{Rec-Movie} & \textbf{Rec-Toy} \\
\cmidrule(lr){2-2} \cmidrule(lr){3-3}
\textbf{Model} & $\mathrm{MRR}$ & $\mathrm{MRR}$ \\
\midrule
Set Encoder              & 6.20 & 1.95 \\
PCA                      & 6.70 & 2.05 \\
Hierarchical Clustering  & \underline{7.00} & \underline{2.18} \\
\midrule
\method                  & \textbf{7.80} & \textbf{2.42} \\
\bottomrule
\end{tabular}}
\end{minipage}
\hfill
\begin{minipage}[h]{0.49\textwidth}
\centering
\resizebox{\linewidth}{!}{
\begin{tabular}{ccc}
\toprule
& \textbf{MS MARCO} & \textbf{ESCI} \\
\cmidrule(lr){2-2} \cmidrule(lr){3-3}
\textbf{Model} & $\mathrm{MRR}$ & $\mathrm{MRR}$ \\
\midrule
Set Encoder              & 43.50 & 50.10 \\
PCA                      & 45.50 & 52.40 \\
Hierarchical Clustering  & \underline{46.80} & \underline{53.80} \\
\midrule
\method                  & \textbf{49.28} & \textbf{57.01} \\
\bottomrule
\end{tabular}}
\end{minipage}

\end{footnotesize}
\end{table*}

\revise{In this section, we compare \method with other methods based on different candidate aggregation methods. To be specific, we design three baselines. To ensure a fair comparison between LRanker and the baselines, we constrain all methods such that the final candidate embeddings fed into the LLM occupy the same number of ''tokens''.}

\begin{itemize}[leftmargin=1em, itemsep=0pt]
    \item \revise{\textbf{Set Encoder}: In this setting, we sample $K$ candidates from the full candidate pool and pass their embeddings through a cross-attention module. The resulting representations are then fed into the LLM. Here, $K$ is set to match the number of k-means cluster centroids used in \method.}  
    
    \item \revise{\textbf{PCA}:  In this setting, the offline embeddings of candidates are first reduced to 256 dimensions using PCA, followed by k-means clustering.}

    \item \revise{\textbf{Hierarchical Clustering}:  Compared with LRanker, in this setting, we replace k-means with hierarchical clustering while keeping the number of clusters unchanged.}
\end{itemize}

\revise{
As shown in Table~\ref{tab:aggregation_ablation}, \method consistently surpasses all three aggregation baselines across all tasks. The Set Encoder performs the worst because it samples only a small subset of candidates and aggregates them through cross attention, inevitably discarding global information from the full candidate pool. The PCA baseline performs better than Set Encoder but still suffers from substantial information loss due to projecting 1024-dimensional embeddings into a 256-dimensional space prior to clustering. Hierarchical Clustering achieves the strongest baseline performance and comes closest to \method, as it preserves more structural relationships and avoids sampling or dimensionality reduction. However, its computational cost is prohibitive: agglomerative hierarchical clustering requires $O(n^{2} d)$ time and $O(n^{2})$ memory to compute and store all pairwise distances, where $n$ is the number of candidates and $d$ is the embedding dimensionality, making it infeasible when $n$ reaches millions. In contrast, k-means used in \method scales as $O(n k d T)$, where $k$ is the number of clusters and $T$ is the number of iterations, thus providing linear rather than quadratic scaling in $n$. As described in Appendix~B, \method further improves the efficiency of k-means by using MiniBatchKMeans, which reduces the effective complexity to $O(b k d T)$ with a small batch size $b \ll n$, and by exploiting the Matryoshka Representation Learning property of Qwen3-Embedding, which enables clustering on truncated embeddings of dimension $d' \ll d$ (128 instead of 1024, as specified in Appendix~B) to reduce clustering cost. These design choices allow \method to maintain strong ranking performance while enabling fast inference under extremely large candidate pools, making it practical for real-world deployment.}

\subsection{\rev{Motivation for Centroid Aggregation}} \label{app:agg_theory}

\rev{We use $k$-means centroids as a compressed summary of the candidate pool. The $k$-means objective, the within-cluster squared error $\sum_{j}\sum_{c_i \in \mathcal{G}_j} \lVert \mathbf{e}(c_i) - \mathbf{g}_j \rVert^2$, is the distortion measure of a $k$-point vector quantizer, which makes centroids a standard way to summarize a set of embeddings with $k$ vectors; Lloyd-style and mini-batch solvers find a local minimum of this objective. Low reconstruction error does not by itself imply better ranking, and we do not claim that centroid aggregation is optimal for ranking or that it preserves the order of candidates within a cluster. The centroids only condition the query embedding; every candidate is still scored with its own embedding (Eq.~(9)).}

\rev{Intuitively, the summary is more faithful when clusters are compact relative to their separation and less faithful when clusters are diffuse or overlap. This intuition is consistent with, but not established by, the empirical results: \method\ stays above the strongest baseline for every tested number of centroids (Figure~\ref{fig:k_sensitivity}), and $k$-means outperforms the alternative aggregators in Table~\ref{tab:aggregation_ablation}, including hierarchical clustering, while scaling as $O(nkdT)$ rather than $O(n^2 d)$. We therefore present centroid aggregation as an empirically supported compression strategy.}

\subsection{\revise{Impact of the Choice of K on Performance}} \label{app:k_choice}

\begin{figure}[ht]
    \centering
    \includegraphics[width=1\linewidth]{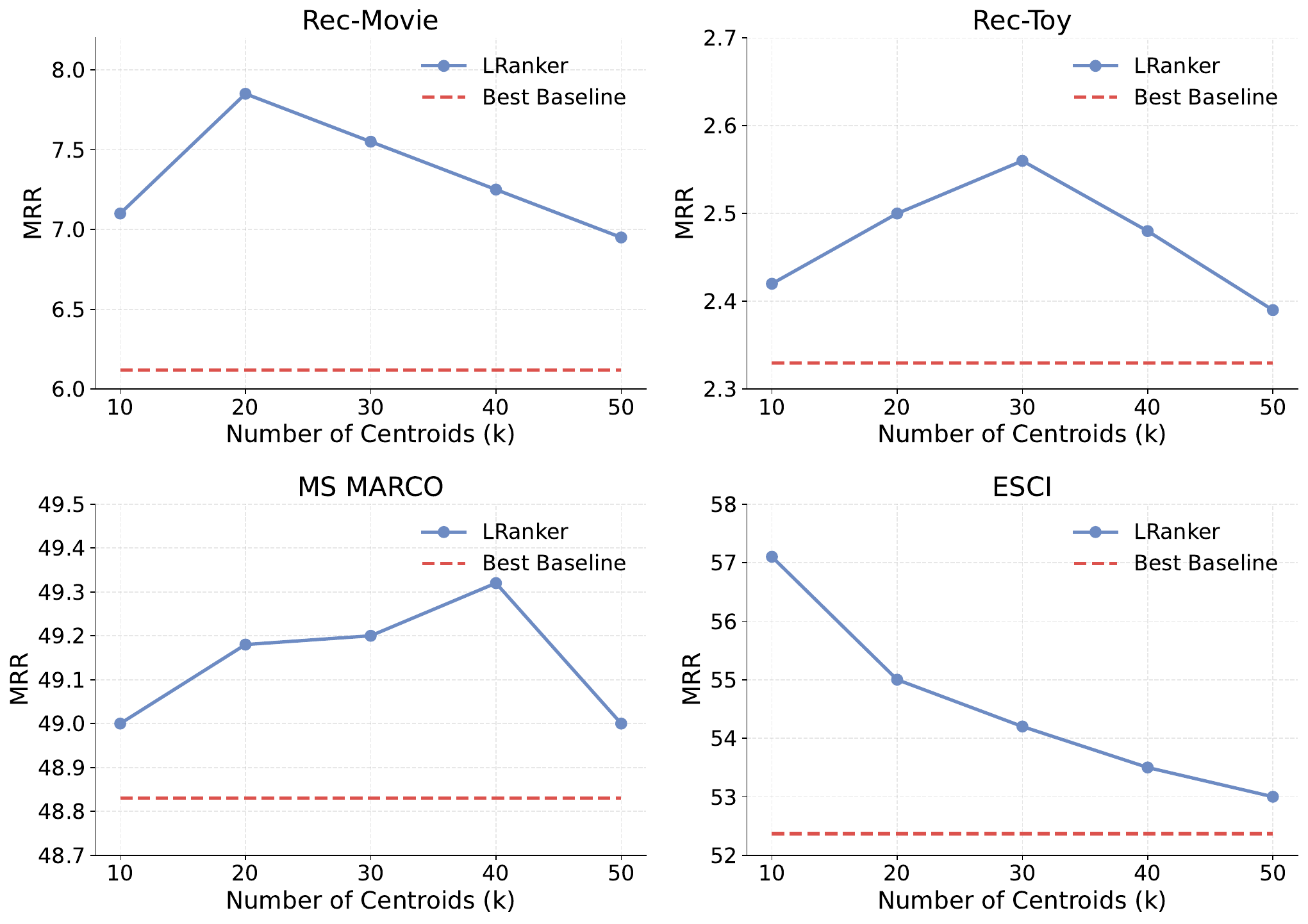}
    \caption{\revise{\textbf{Effect of the number of centroids ($K$) on MRR across four tasks.} The horizontal-axis label $k$ in the plot denotes centroid count $K$, not test-time partition width. All plotted settings exceed the displayed baseline. The best tested centroid count is task-dependent: 20 for Rec-Movie, 30 for Rec-Toy, 40 for MS MARCO, and 10 for ESCI.}}
\label{fig:k_sensitivity}
\end{figure}

\revise{We vary the number of centroids $K$ over $\{10,20,30,40,50\}$ (Figure~\ref{fig:k_sensitivity}). These are the five tested values, rather than an evaluation of every integer in the interval. The best plotted value is $K=20$ for Rec-Movie, $K=30$ for Rec-Toy, $K=40$ for MS MARCO, and $K=10$ for ESCI. Performance is therefore task-dependent and not monotonic in centroid count. All plotted configurations remain above the displayed baseline. The experiment concerns centroid count, which is distinct from the partition width $k$ used in test-time scaling.}

\subsection{\revise{Effect of Centroid Dimensionality on Model Performance}} \label{app:dimension}

\begin{figure}[ht]
    \centering
    \includegraphics[width=1\linewidth]{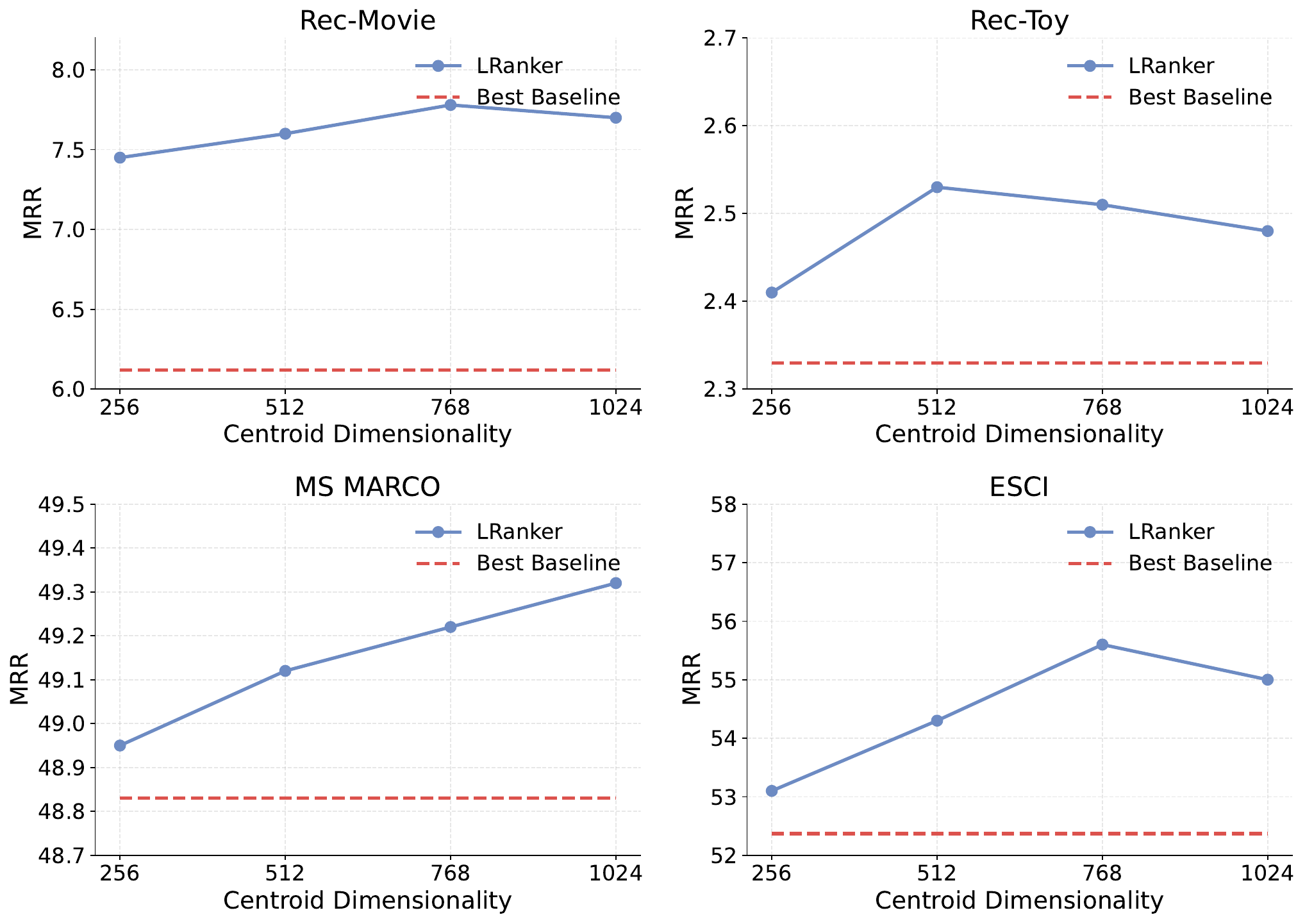}
    \caption{\revise{\textbf{Effect of centroid dimensionality on MRR across four tasks.} Performance is not uniformly monotonic: Rec-Movie and ESCI peak at 768 dimensions, Rec-Toy at 512, and MS MARCO at 1024 among the tested settings. All plotted settings exceed the displayed baseline.}}
\label{fig:dimension_sensitivity}
\end{figure}

\revise{We vary centroid dimensionality over $\{256,512,768,1024\}$ (Figure~\ref{fig:dimension_sensitivity}). Among these settings, Rec-Movie and ESCI peak at 768 dimensions, Rec-Toy at 512, and MS MARCO at 1024. Thus, increasing dimensionality does not uniformly improve MRR, and the best tested dimensionality is task-dependent. This experiment varies the retained centroid representation; it is distinct from the 128-dimensional vectors used to compute cluster assignments in Appendix~B.}

\subsection{\revise{Analysis of Backbone LLM Influence}} \label{app:backbone}

\begin{table*}[ht]
\setlength\tabcolsep{3pt}
\centering
\begin{footnotesize}
\vspace{-6mm}
\caption{\revise{\textbf{Evaluating the direct task-solving capability of Qwen3 backbones of different sizes against general and task-specific baselines on Rec-Movie and Rec-Toy.} \textbf{Bold} and \underline{underline} denote the best and second-best results.}}
\label{tab:back}
\centering
\resizebox{0.65\textwidth}{!}{
\begin{tabular}{lcccc}
\toprule
& \multicolumn{2}{c}{\textbf{Rec-Movie}} & \multicolumn{2}{c}{\textbf{Rec-Toy}} \\
\cmidrule(lr){2-3} \cmidrule(lr){4-5}
\textbf{Model} & $\mathrm{NDCG}@10$ & $\mathrm{MRR}$ & $\mathrm{NDCG}@10$ & $\mathrm{MRR}$ \\
\midrule
\multicolumn{1}{c}{\textbf{\textit{General Ranking Baselines}}} \\
\midrule
BM25       & 0.18 & 0.54 & 0.37 & 0.42 \\
Contriever & 0.24 & 0.43 & 0.84 & 1.11 \\
\midrule
\multicolumn{1}{c}{\textbf{\textit{Backbone Models}}} \\
\midrule
Qwen3 0.6B       & 0.61 & 0.84 & 1.56 & 1.52 \\
Qwen3 4B & 0.84 & 1.29 & 2.52 & 2.11 \\
\midrule
\multicolumn{1}{c}{\textbf{\textit{Task-specific Baselines}}} \\
\midrule
FM     & 2.35 & 2.01 & 0.95 & 0.98 \\
BERT4Rec     & 4.08 & 3.56 & 1.26 & 1.31 \\
GRU4Rec     & 4.12 & 3.59 & 1.59 & 1.46 \\
SASRec     & 4.36 & 3.84 & 1.65 & 1.52 \\
Tiger     & \underline{7.37} & \underline{6.12} & \underline{2.99} & \underline{2.33} \\
\midrule
\midrule
\method & \textbf{8.02} & \textbf{7.80} & \textbf{3.21} & \textbf{2.42} \\
\bottomrule
\end{tabular}}
\end{footnotesize}
\end{table*}

\revise{In addition to comparing against general and task-specific baselines, we further
evaluate the ranking capability of the backbone models used in \method by directly
applying two Qwen3 variants (0.6B and 4B) to the Rec-Movie and Rec-Toy tasks in a
zero-shot manner. As shown in Table~\ref{tab:back}, the 0.6B model
performs substantially worse than most task-specific baselines, indicating that a
small LLM backbone lacks sufficient inductive bias for effective item ranking. The
larger 4B model shows noticeable improvement, yet still lags behind the strongest
baselines (e.g., Tiger), demonstrating that simply scaling the backbone model size
does not close the gap. These results highlight that directly using a pretrained
Qwen3 model to solve ranking tasks does not inherently provide performance gains.
In contrast, \method achieves substantial improvements by combining a carefully
designed training objective with graph-based test-time scaling, which equips the
LLM with candidate aggregation encoder and enables robust ranking behavior
far beyond what the backbone alone can offer.}

\section{\revise{Computation and runtime  analysis}} \label{app:com}

\subsection{\revise{Evaluation of Computational Cost and Runtime Against Best Baselines}}
\label{app:best_baselines}

\begin{table}[ht]
\centering
\caption{\revise{Computation time and memory usage for different models on Rec-Movie and MS MARCO.}}
\label{tab:computation_cost}
\begin{footnotesize}
\begin{tabular}{lcccccc}
\toprule
\textbf{Scenario} & \textbf{Model} & \textbf{Train Time} & \textbf{Train Memory} & \textbf{Test Time} & \textbf{Test Memory} \\
\midrule
Rec-Movie 
  & Tiger        & 21 h 36 m  & 10.8 GB  & 20 min    & 200 MB   \\
  & LRanker      & 52 min     & 24.5 GB  & 15 min    & 17.2 GB  \\
\midrule
MS MARCO 
  & RankLLaMA 8B & 6 h 33 min & 188 GB   & 21.67 min & 71.28 GB \\
  & LRanker      & 21 min     & 25.0 GB  & 10 min    & 17.5 GB  \\
\bottomrule
\end{tabular}
\end{footnotesize}
\end{table}

\revise{As shown in Table~\ref{tab:computation_cost}, we compare the computation time and memory usage of LRanker against the strongest task-specific baselines on Rec-Movie and MS~MARCO under identical hardware settings. On Rec-Movie, LRanker completes training in only 52 minutes, representing more than a \emph{24$\times$ reduction} in training time compared with Tiger (21 h 36 m), while also achieving lower test-time latency (15 min vs.\ 20 min). A similar trend appears on MS~MARCO: LRanker requires just 21 minutes to train, in stark contrast to RankLLaMA~8B, which takes 6 h 33 m. Test-time latency is also reduced by more than half (10 min vs.\ 21.67 min).  The memory trade-off differs by baseline: LRanker uses more training and inference memory than Tiger on Rec-Movie (24.5 GB vs.\ 10.8 GB; 17.2 GB vs.\ 200 MB), but less than RankLLaMA on MS MARCO (25.0 GB vs.\ 188 GB; 17.5 GB vs.\ 71.28 GB). These results highlight that LRanker achieves substantial improvements in computational efficiency and latency without sacrificing effectiveness. Overall, LRanker provides a practical and scalable solution for real-world retrieval and ranking systems, where fast training and low-latency inference are essential.}

\subsection{\revise{Scenario-Wise Memory Usage of \method}} 
\label{app:all_sce}

\begin{table}[ht]
\centering
\caption{\revise{Memory Requirements of RBench Scenarios.}}
\label{tab:memory_requirements}
\begin{tabular}{lcc}
\toprule
\textbf{Scenario} & \textbf{Train Memory} & \textbf{Inference Memory} \\
\midrule
Rec-Music & 24.4 GB & 16.9 GB \\
Routing-Balance & 24.4 GB & 16.9 GB \\
Rec-Movie & 24.5 GB & 17.2 GB \\
ESCI & 24.5 GB & 17.2 GB \\
Rec-Toy & 25.0 GB & 17.5 GB \\
MS MARCO & 25.0 GB & 17.5 GB \\
Rec-Clothing & 37.2 GB & 29.2 GB \\
\bottomrule
\end{tabular}
\end{table}

\revise{Table~\ref{tab:memory_requirements} reports training and inference memory for seven RBench tasks. The six non-million-scale tasks use 24.4--25.0 GB for training and 16.9--17.5 GB for inference. Rec-Clothing uses more memory: 37.2 GB and 29.2 GB, respectively. These measurements do not imply memory usage independent of candidate count, and the table does not report memory for the 48M-candidate extension.}

\subsection{\revise{Latency comparison in Table 1}}
\label{app:latency}

\begin{table}[ht]
\centering
\caption{\textbf{\revise{Per-query inference latency comparison across different models on Rec-Music and MS MARCO.}} \rev{Reported times are the \emph{online}, per-query cost. For \method, the time is for direct ranking (Step~1 in Section~\ref{sec:tts}): projecting the centroids, one LLM forward pass, inner-product scoring of all candidates with their cached embeddings (Eq.~(9)), and sorting; the added latency of test-time scaling is reported in Table~\ref{tab:width_depth}. The base-encoder candidate embedding pass, the candidate embeddings $\mathbf{h}_{c_i}$ used for scoring, and the $K$-means centroid construction are one-time \emph{offline} preprocessing that is amortized across all queries and is not included. For the baselines, the time covers each method's full per-query ranking procedure. The compared methods span different operating regimes---several baselines are not designed for million-scale pools---so this table characterizes the asymptotic scaling gap rather than a same-regime head-to-head.}}
\label{tab:inference_latency}
\begin{footnotesize}
\begin{tabular}{lcc}
\toprule
\textbf{Scenario} & \textbf{Model} & \textbf{\rev{Online} Per-query Inference Time} \\
\midrule
Rec-Music (20 candidates)
    & PRP           & 38.3 s \\
    & IRanker       & 6.5 s  \\
    & RankGPT       & 3.8 s  \\
    & LRanker       & 15 ms  \\
\midrule
MS MARCO (24{,}697 candidates)
    & RankLLaMA 8B  & 21.67 min \\
    & LRanker       & 20 ms      \\
\bottomrule
\end{tabular}
\end{footnotesize}
\end{table}

\revise{Table~\ref{tab:inference_latency} shows that \method delivers consistently low per-query latency across both small and large candidate pools. On Rec-Music (20 candidates), models such as PRP, IRanker, and RankGPT require seconds of computation, while \method responds in only 15\,ms. The gap widens on MS~MARCO, where RankLLaMA~8B needs over 21 minutes per query due to full-candidate scoring, whereas \method maintains a 20\,ms latency \rev{because its LLM reads precomputed centroids rather than candidate text, while every candidate is still scored with a cached embedding (Eq.~(9))}. These are reported measurements for the listed configurations; they do not establish production latency or performance at million-scale candidate counts.} \rev{To be precise about the source of this efficiency, we distinguish prompt length from total computation. For a fixed number of centroids $K$, the \emph{prompt length} is independent of the candidate count $N$, which is the sense in which Table~\ref{Tab:intro} describes \method's input prompt as independent of candidate size; for a fixed width $k$ and depth $d$, so is the number of LLM forward passes ($1 + d\cdot k$; the selected $k$ and $d$ differ by task, Table~\ref{tab:width_depth}). Total computation still grows with $N$. Online, \method\ projects the centroids, runs the LLM forward pass(es), scores all $N$ candidates by inner product with their cached embeddings (Eq.~(9)), and sorts them; with test-time scaling it also prunes candidates and recomputes per-partition centroids in each round. The 15--20\,ms in Table~\ref{tab:inference_latency} cover direct ranking (projection, one forward pass, scoring, and sorting), and Table~\ref{tab:width_depth} reports the additional latency of test-time scaling. Offline, \method\ encodes all $N$ candidates (both $\mathbf{e}(c_i)$ and $\mathbf{h}_{c_i}$), runs MiniBatchKMeans at a cost of $O(bKDT)$ for mini-batch size $b$, $K$ centroids, embedding dimension $D$, and $T$ iterations, and assigns all $N$ candidates to their clusters (Appendix~B); these steps are linear in $N$ and are amortized across all subsequent queries.}

\subsection{\revise{Impact of  Widths and Depths on the Computational Efficiency of \method}}
\label{app:width_depth}

\begin{table}[ht]
\centering
\caption{\textbf{\revise{Optimal width/depth settings of LRanker and their relative latency overhead compared to direct ranking.}} \rev{Direct ranking is the setting timed in Table~\ref{tab:inference_latency} (one LLM forward pass on the full candidate set, scoring, and sorting). The latency increase is the additional per-query latency of the $d$ test-time scaling rounds, each with $k$ partition encodings, per-partition centroid recomputation, and pruning.}}
\label{tab:width_depth}
\begin{tabular}{lccc}
\toprule
\textbf{Scenario} & \textbf{Best Width} & \textbf{Best Depth} & \textbf{Latency Increase vs. Direct Ranking (\%)} \\
\midrule
\multicolumn{4}{c}{\textbf{RBench-Small}} \\
\midrule
Rec-Music        & 3  & 3 & 1.2\% \\
Routing-Balance  & 3  & 3 & 1.5\% \\
\midrule
\multicolumn{4}{c}{\textbf{RBench-Large}} \\
\midrule
Rec-Movie        & 5  & 4 & 3.2\% \\
Rec-Toy          & 5  & 6 & 3.8\% \\
MS MARCO         & 9  & 6 & 6.5\% \\
ESCI             & 6  & 5 & 5.2\% \\
\midrule
\multicolumn{4}{c}{\textbf{RBench-Ultra}} \\
\midrule
Rec-Clothing     & 10 & 6 & 7.8\% \\
\bottomrule
\end{tabular}
\end{table}

\revise{As shown in Table~\ref{tab:width_depth}, the additional latency introduced by the
graph-based test-time scaling module is surprisingly small across all scenarios.
Even when the best-performing configurations require moderate widths and depths
(e.g., width $=9$, depth $=6$ on MS~MARCO), the relative latency increase over
direct ranking remains below $8\%$, and is often as low as $1$--$3\%$ on the
smaller RBench tasks. This demonstrates that LRanker achieves substantial ranking
improvements with minimal overhead. Moreover, when viewed in absolute terms, LRanker remains extremely fast.  As reported in Appendix~F.3, its per-query inference latency is already orders of
magnitude lower than that of strong LLM-based rankers (e.g., RankLLaMA~8B),
and remains competitive even against lightweight models such as RankGPT.  
This high efficiency is enabled by the test-time design described in Appendix~B:
MiniBatchKMeans for scalable centroid construction, Matryoshka Representation
Learning (MRL) for low-dimensional clustering, and centroid-only aggregation for
compact LLM input. Together, these optimizations ensure that LRanker maintains
both strong performance and low latency, even when operating under larger
width/depth configurations.}

\section{\rev{Broader Impact}} \label{app:impact}

\rev{\method\ improves the efficiency and scalability of LLM-based ranking, which can reduce the computational cost of large-scale search and recommendation. Two considerations merit attention. First, because the global summary is formed by averaging candidate embeddings into centroids, the aggregated representation reflects the dominant structure of the candidate pool; in applications where the pool is already skewed toward popular content, deployments should ensure that this does not amplify popularity bias. Importantly, \method's final decision is not made by the aggregated vector: every candidate is scored by an explicit, inspectable inner product $\langle \mathbf{h}_q, \mathbf{h}_{c_i} \rangle$ (Eq.~(9)), and the elimination step operates on these query-conditioned scores rather than on item popularity directly, although popularity can still enter through the learned scores, and we have not evaluated long-tail or cold-start performance (Appendix~\ref{app:longtail}). Second, the injected soft prompt $\tilde{\mathbf{g}}$ is a continuous vector rather than human-readable text, which reduces the direct interpretability of the conditioning signal; the transparency of the per-candidate scoring function nonetheless supports post-hoc inspection of why a candidate was ranked highly. In high-stakes retrieval settings such as hiring, legal, or medical search, we recommend pairing \method\ with domain-appropriate auditing of ranked outputs.}

\section{LLM Writing Usage Disclosure} \label{app:LLM usage}

An LLM was applied as a writing aid to enhance the clarity and linguistic quality of this paper, specifically by correcting grammatical errors and polishing sentence flow. No part of the research design, data analysis, or interpretation relied on the use of the LLM.

\end{document}